\documentclass[aps,superscriptaddress,showpacs,nofootinbib,floatfix,epfs,superscriptaddress,showpacs,groupedaddress,preprintnumbers]{revtex4}
\usepackage{amsfonts,amscd,amsmath,amssymb,graphicx,color,relsize}
\usepackage[section]{placeins}
\def\vp{v}
\def\rv{r_\text{\tiny v}}
\def\fv{f_\text{\tiny v}}
\def\est{\sigma_\text{\tiny eff}}
\def\rmin{r_\text{\tiny min}}
\def\rmax{r_\text{\tiny max}}
\def\lmax{\ell_\text{\tiny max}}
\def\qqb{\text{\tiny Q}\bar{\text{\tiny Q}}}
\def\qu{\text{\tiny Q}}
\def\aqu{\bar{\text{\tiny Q}}}
\def\rh{r_h}

\def\et2q{\eta_{\text{\tiny QQ}}}

\def\D2q{D_{\text{\tiny QQ}}}

\def\rv{r_v}

\def\Td{T_{\text{\scriptsize diss}}}

\def\h0{h_0}

\def\rmax{r_{\text{\tiny max}}}

\def\erfi{\text{Erfi}}
\def\oh{\frac{1}{2}}

\def\r0{r_{\text{\tiny 0}}}
\def\ep{\text{e}}
\def\g{\mathfrak{g}}
\def\oh{\frac{1}{2}}

\def\s{\mathfrak{s}}

\begin{document}
\title{Aspects of quarkonium propagation in a thermal medium as seen by string models}
\author{Oleg Andreev}
 \affiliation{L.D. Landau Institute for Theoretical Physics, Kosygina 2, 119334 Moscow, Russia}
\affiliation{Arnold Sommerfeld Center for Theoretical Physics, LMU-M\"unchen, Theresienstrasse 37, 80333 M\"unchen, Germany}
\begin{abstract} 
We use gauge/string duality to model a heavy quark-antiquark pair in a color singlet moving through a thermal plasma. In particular, we explore the effect of velocity on the string tension and Debye screening mass. Then we apply the results to the analysis of heavy quarkonium bound states. With some assumptions, we estimate the characteristic size of quarkonium and its dissociation temperature. 

\pacs{11.25.Tq, 12.38.Lg, 12.38.Mh}
\preprint{LMU-ASC 05/19}
\end{abstract}
\maketitle


\section{Introduction}

Heavy-ion collision experiments being performed at BNL and CERN aim to recreate a state of matter known as the quark-gluon plasma (QGP). In this regard, the question arises: how to probe the properties of the plasma formed in such collisions? One of the ideas put forward by Matsui and Satz \cite{satz} is that quarkonium production rates, in particular that of the $J/\Psi$ mesons, are a good indicator of whether the QGP is produced, and at what temperature.\footnote{The literature on quarkonium suppression in   heavy-ion collisions is vast. For a review, see, e.g., \cite{bram-rev,uaw-book,satz-book} and references therein.} In this case the understanding of the in-medium behavior of heavy quark bound states is the main ingredient of the analysis with the key feature that the QGP formation leads to a color screening and, as a consequence, no existence of bound states at high enough temperature. 

In reality, the physics is rather complicated, and as yet there is no complete understanding of quarkonium suppression in heavy-ion collisions. To be more accurate, the analysis has to include various confounding effects \cite{uaw-book}. In this paper, we will focus on one which has a particularly simple origin: quarkonium mesons may be produced moving through a thermal medium. Then the question is what happens to the suppression of those mesons? 

Since the original paper in 1986, the lattice calculations of color screening \cite{lattice-rev} have played a pivotal role in the analysis of bound states. Although lattice QCD is a powerful computational tool to deal with strongly coupled gauge theories, its use is limited to the static case when heavy quark bound states are at rest. Meanwhile the gauge/string duality allows one to at least get a good intuition in situations where evolution in real time is an essential feature, as in the case of interest. This makes it an effective tool to tackle the problem. It is worth noting that such a duality was first proposed by Maldacena for conformal theories and then expanded to include QCD-like theories, and thus somehow relate string theory to heavy-ion collisions \cite{uaw-book,ssg}. Because there still is no string theory dual to QCD, effective string models will only be used for our purposes.

The present paper continues a series of studies on five-dimensional string models started in \cite{az3,a-screen}. For orientation, we begin by describing a thermal medium from the viewpoint of a black hole in five dimensions. We go on in section III to discuss general aspects of a quark-antiquark pair propagating in the medium. In section IV, we analyze two important quantities for characterizing the pair: the string tension and Debye screening mass. In doing so, we particularly focus on the velocity dependence of those. Then, in section V, we use the results to make two phenomenological estimates. These include the estimates of the characteristic size of quarkonium and its dissociation temperature in the presence of a thermal wind. Finally, we conclude in section VI. Some technical details are given in the appendices.

\section{Preliminaries}
\renewcommand{\theequation}{2.\arabic{equation}}
\setcounter{equation}{0}

As a preliminary to discussing a heavy quark pair propagating in a thermal medium, we will first provide some background information on the medium itself. 

It is well known that the thermodynamics of $N = 4$ super Yang-Mills theory in four dimensions is related to the thermodynamics of the Schwarzschild black hole in a five-dimensional Anti-de Sitter space \cite{wittenT}. For QCD-like theories the situation is more involved, as the geometries become much complicated and often include additional background fields. In this paper we will discuss a pair of {\it effective} string models. Our reasons for doing so are: (1) Because it would be very good to gain some intuition on problems in QCD that can be done within the effective string models already at our disposal. (2) Because the string theory dual to QCD remains still unknown. (3) Because the results provided by the models are consistent with the lattice calculations and QCD phenomenology. In particular, this takes place for the equation of state \cite{a-T2}, expectation value of the Polyakov loop \cite{a-pol}, Debye screening mass \cite{a-screen}, spatial string tension and heavy quark diffusion coefficients \cite{a-D}. (4) Because analytic formulas can be obtained by solving these models. (5) Because the aim of our work is to make predictions which may then be tested by means of other methods, e.g., effective field theories or numerically.

The five-dimensional metric (with Euclidean signature) is taken to be

\begin{equation}\label{metric}
ds^2=w(r)R^2\Bigl(f(r)dt^2+dx_i^2+f^{-1}(r)dr^2\Bigr)
\,,
\quad\text{with}
\quad
w(r)=\frac{\ep^{\s r^2}}{r^2}
\,.
\end{equation}
We assume that the blackening factor $f$ is a decreasing function of $r$ on the interval $[0,\rh]$ such that $f(0)=1$ at the boundary and $f(\rh)=0$ at the horizon. $t$ is a periodic variable of period $1/T$. The index $i$ goes from $1$ to $3$. The boundary of this space, thus, is $\mathbf{R}^3\times\mathbf{S}^1$. Note that one can think of this geometry as a one-parameter deformation, parameterized by $\s$, of the Schwarzschild black hole on $\text{AdS}_5$ space of radius $R$. 

Two forms of $f$ have been discussed in the present context. One, as originally postulated in \cite{az2}\footnote{Actually, this was the first attempt to extend the soft wall (metric) model \cite{soft} to finite temperature.}, is given by 

\begin{equation}\label{f-az2}
f(r)=1-\Bigl(\frac{r}{\rh}\Bigr)^4
\,,
\end{equation}
which is the blackening factor of the Schwarzschild black hole in $\text{AdS}_5$ space. With this choice, the Hawking temperature is simply

\begin{equation}\label{Taz2}
T=\frac{1}{\pi}\sqrt{\frac{\s}{h}}
\,,
\end{equation}
where $h = \s\rh^2$. This is the simplest case that enables one to write most of the resulting equations analytically. 

Another one was derived from a differential equation which relates the functions $w(r)$ and $f(r)$. It follows from the Einstein equations and can be solved explicitly \cite{noro}. For $w$ given by \eqref{metric}, one gets 

\begin{equation}\label{f-noro}
f(r)=1-
\frac{1-\bigl(1+\tfrac{3}{2}\s r^2\bigr)\,\ep^{-\tfrac{3}{2}\s r^2}}
{1-\bigl(1+\tfrac{3}{2}h\bigr)\,\ep^{-\tfrac{3}{2}h}}
\,.
\end{equation}
Given this, the corresponding temperature is 

\begin{equation}\label{Tnoro}
T(h)=\frac{9}{8\pi}
\frac{\sqrt{\s}\,h^{\frac{3}{2}}}{\ep^{\tfrac{3}{2}h}-1-\tfrac{3}{2}h}
\,.
\end{equation}
These blackening factors lead to results which coincide with each other at high temperatures, but display some differences at lower temperatures, especially near the critical point.

We will use both backgrounds in our analysis, suggesting that the worldsheet path integral for the expectation value of Polyakov loops can be evaluated on some classical solutions which describe the minimal surfaces. At that point there is no requirement on the background metric to be a solution of $\beta_{\mu\nu}=0$, with $\beta_{\mu\nu}$ the worldsheet beta function \cite{gsw}. Of course, such an approximation requires a caveat, because of quantum corrections (effects of order $\alpha'$).  
\section{A moving quark-antiquark pair }
\renewcommand{\theequation}{3.\arabic{equation}}
\setcounter{equation}{0}

Now we turn to a quark-antiquark pair moving thought a thermal medium. In doing so, we assume that the quarks are heavy enough such that there is no relative motion between them. It is also worth keeping in mind one fact: in the dual formulation we are using the pair does not feel a drag force and, consequently, does not lose its energy \cite{arg}. In other words, a string configuration is moving uniformly through the medium. Of course, it is an assumption. In the reality there is energy lost that makes string configurations time dependent and, as a consequence, the analysis becomes extremely tedious and complicated. 

In terms of strings in five dimensions, the pair is described by a Nambu-Goto string attached to quark sources on the boundary. We place the quarks on the $x$-axis and then consider their uniform motion with velocity $v$ in the $z$-direction, as sketched in Figure \ref{configs}. The action is the sum of the string action and those of two point-like particles. For our purposes in this paper, what we need to know is how the energy of this system depends on the quark separation. This can be determined entirely from the Nambu-Goto action 

\begin{equation}\label{NG}
S=-\frac{1}{2\pi\alpha'}\int d^2\xi\,\sqrt{-\det\gamma_{\alpha\beta}}
\,,
\end{equation}
where $\alpha'$ is a string constant, $(\xi_1,\xi_2)$ are worldsheet coordinates, and $\gamma_{\alpha\beta}$ is an induced metric. 

In studying bound state properties, it is more convenient to choose the rest frame of the pair. In that case, the medium takes on the role of a hot wind blowing on the static pair. To describe such a wind geometrically, one first Wick rotates to Minkowski space and then boosts the metric in the $z$-direction \cite{uaw-wind}. With $t=\gamma(t'+vz')$ and $z=\gamma(z'+vt')$, a simple algebra shows that the metric \eqref{metric} takes the form\footnote{To avoid clutter, we suppress the primes as we will always be interested in the pair rest (primed) frame.}   

\begin{equation}\label{metricv}
ds^2=w(r)R^2
\Bigl(-\fv dt^2+2\vp \gamma^2(1-f)dt dz
+
dx^2+dy^2+\gamma^2(1-f\vp^2)dz^2
+
 f^{-1}dr^2\Bigr)
\,,
\end{equation}
where $\gamma=1/\sqrt{1-v^2}$. For convenience, we introduce a new blackening factor 

\begin{equation}\label{fv}
\fv(r)=\gamma^2(f-\vp^2)
\,.
\end{equation}
This metric describes the hot wind felt by the static pair positioned at a right angle to the direction of the wind.

\subsection{A connected string configuration}

We now want to consider a connected string configuration. It is similar to the configuration sketched in Figure \ref{configs}, but with $\rh$ replaced by $\rv$. The analysis proceeds along the lines described in \cite{az3}.

We start by taking the static gauge $t=\xi_1$ and $x=a\,\xi_2+b$. Combining the latter with the boundary conditions at the string endpoints $x(0)=-\frac{\ell}{2}$ and $x(1)=\frac{\ell}{2}$, we obtain $a=\ell$ and $b=-\frac{\ell}{2}$. Since the configuration is static, the lowest energy corresponds to $z=const$, and therefore $r$ is a function of $x$ only, as in the $\text{AdS}$ case \cite{chernic}. For such a configuration, the induced metric is given by 

\begin{equation}\label{metric-induced}
ds^2=w(r)R^2
\Bigl(-\fv dt^2+(1+f^{-1}r'{}^2)dx^2\Bigr)
\,.
\end{equation}
This metric has a horizon at $r=\rv$, where the equation $f(r)=v^2$ has a solution. We call it an induced horizon as a shorthand for saying that it is related to the induced metric on the string worldsheet. Note that, since $f$ is a decreasing function of $r$ such that $f(\rh)=0$, it holds $\rv<\rh$ for all $v\not =0$. Also, it is useful for further analysis to introduce the notion of an effective string tension 

\begin{equation}\label{est}
	\est(r)=\frac{\ep^{\s r^2}}{r^2}\sqrt{\fv}
	\,,
\end{equation} 
and think of it as a function of $r$.

The Nambu-Goto action now becomes

\begin{equation}\label{NG-QQ}
	S=-\g\int dt \int_{-\ell/2}^{\ell/2} dx\,
	\est\sqrt{1+f^{-1}r'{}^2}
	\,,
	\end{equation}
where $\g=\frac{R^2}{2\pi\alpha'}$ and $r'$ means $\frac{\partial r}{\partial x}$. Since the Lagrangian does not depend on $x$ explicitly, there is a conserved quantity. It is nothing else but the first integral of the equation of motion for $r$ 

\begin{equation}\label{I}
I=\frac{\est}{\sqrt{1+f^{-1}r'{}^2}}
\,.
\end{equation}
On symmetry grounds, $r'(0)=0$ that allows one to express $I$ in terms of $\est$ as $I=\est(\r0)$, where $r(0)=\r0$. 

The string length along the $x$-axis can be found by further integrating \eqref{I}. Writing it as a relation between $dx$ and $dr$ and then performing the integral over $[-\frac{\ell}{2},\frac{\ell}{2}]$ over $x$, we obtain

\begin{equation}\label{l}
\ell=2\int_0^{\r0} \frac{dr}{\sqrt{f}}\, 
\biggl[\,
\frac{\est^2}{I^2}-1\,
\biggr]^{-\oh}
\,,
\end{equation}
where the factor $2$ comes from the reflectional symmetry of the configuration.

Given a string configuration (solution), we can compute its energy. Since the string is static, the energy is simply related to the Lagrangian. So, 

\begin{equation}\label{E}
E_{\qqb}=2\g\int_0^{\r0} \frac{dr}{\sqrt{f}}\, \est
\biggl[\,1-\frac{I^2}{\est^2}\,\biggr]^{-\oh}
\,.
\end{equation}
Here we used \eqref{I} to reduce the integral over $x$ to an integral over $r$. This expression is not well-defined, because the integral diverges at $r=0$. We regularize it by imposing a cutoff $\epsilon$ such that $r\geq\epsilon$. Then in the limit $\epsilon\rightarrow 0$ the regularized expression behaves like 

\begin{equation}\label{ER}
E_{\qqb}^{\text{\tiny R}}=\frac{2\g}{\epsilon}+E_{\qqb}+O(\epsilon)
\,.
\end{equation}
Subtracting the $\frac{1}{\epsilon}$ term and letting $\epsilon=0$, we get a finite result

\begin{equation}\label{E2}
E_{\qqb}=2\g\int_{0}^{\r0} dr\biggl(\frac{\est}{\sqrt{f}}
\biggl[\,1-\frac{I^2}{\est^2}\,\biggr]^{-\oh}-\frac{1}{r^2}\biggr)\,\,-\frac{2\g}{\r0}+2c
\,,
\end{equation}
where $c$ is a normalization constant. Thus, the energy of the string is given by the parametric equations \eqref{l} and \eqref{E2}, with $\r0$ a parameter.\footnote{When written in this form, the formulas are also applicable to other backgrounds, like those with $w=\frac{\ep^{A(r)}}{r^2}$.} It can take values in the interval $[0,\rv]$. The reason for this is that both $E_{\qqb}$ and $E_\qu$ (as defined below) must be real-valued. As an immediate comment, we note that for $v=0$, the formulas are reduced to those of \cite{az3} obtained for the string free energy. 

It is now straightforward to investigate the properties of $E_{\qqb}$ at short distances. In doing so, an important fact will be that small $\ell$'s correspond to small $\r0$'s. This means that a short string stays close to the boundary. The effective string tension and blackening factor behave for $r\rightarrow 0$ as 
$\est\sim 1/r^2$ and $f\sim 1$. Therefore, in this limit, the behavior of $E_{\qqb}$ is the same as in the $\text{AdS}$ case at zero temperature and velocity \cite{malda}. We have

\begin{equation}\label{El-small}
	E_{\qqb}=-\frac{\alpha_{\qqb}}{\ell}+2c+o(1)
	\,,
\end{equation}
with $\alpha_{\qqb}=\g\frac{(2\pi)^3}{\Gamma^{4}\bigl(\tfrac{1}{4}\bigr)}$, and this holds for a large class of backgrounds.

\begin{figure*}[htbp]
\centering
\includegraphics[width=6.75cm]{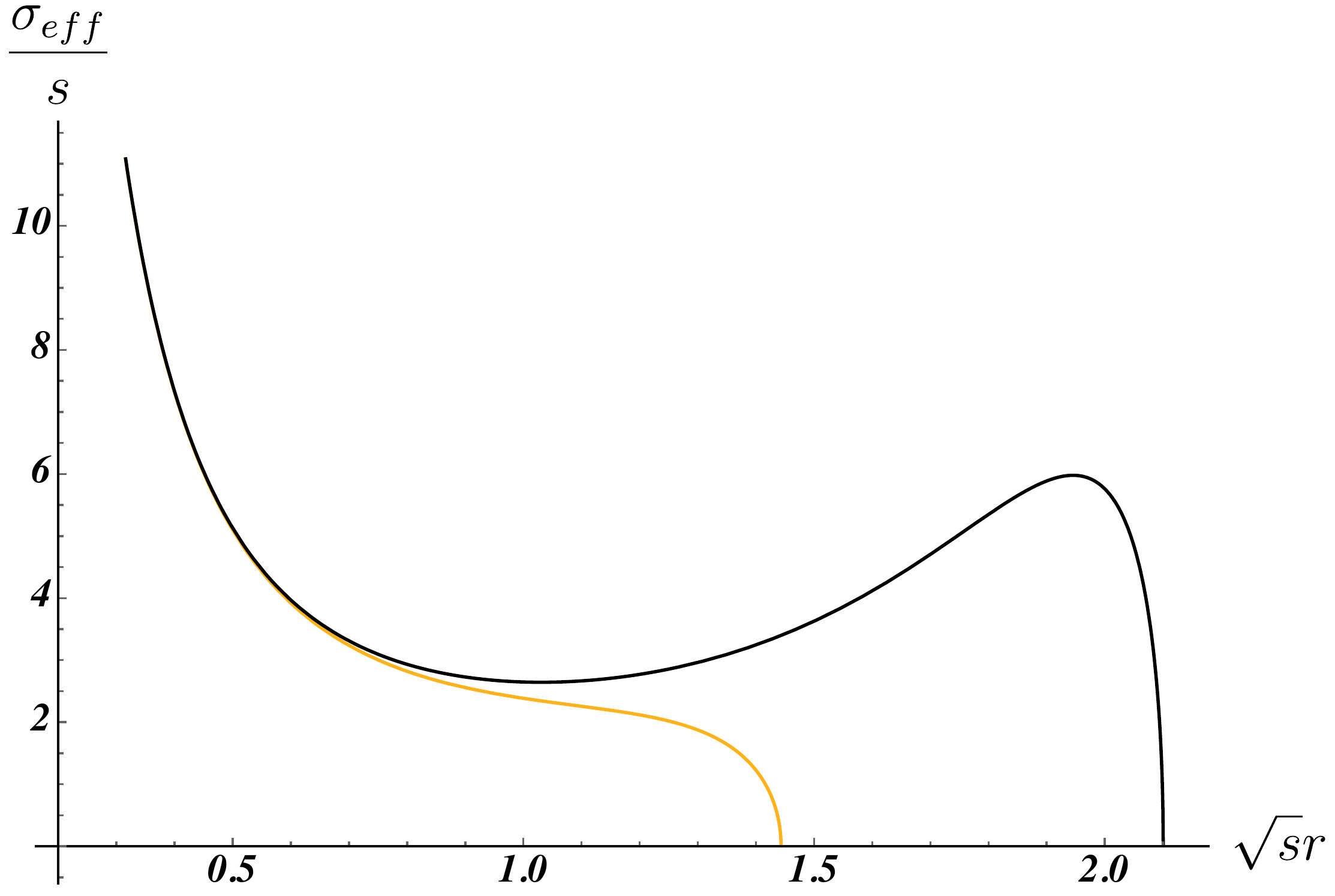}
\caption{{\small Schematic representation of the effective string tension in two different regimes: confinement (upper curve) and deconfinement (lower curve).}}
\label{Est}
\end{figure*}

Let us now investigate the long distance behavior of $E_{\qqb}$ that is achievable with a little more effort. Following \cite{az3}, we start with the effective string tension. There is one very simple but important fact: the form of $\est$, as a function of $r$, is temperature and velocity dependent.\footnote{For the Reissner-Nordstr\"om blackhole whose blackening factor reduces to that of \cite{az2} at zero chemical potential, it was recently discussed in \cite{ch}.} At the qualitative level, it can be understood as follows. For small $T$ and $v$, where $f\approx 1$, $\est$ has a local minimum near $r=1/\sqrt{\s}$ defined by the warp factor. For large enough $T$ and $v$, where $\rv\ll 1/\sqrt{\s}$, there is no local minimum as both factors are decreasing functions in the interval $[0,\rv]$. This is illustrated in Figure \ref{Est}.
We return to this issue in Section IV, after describing the behavior of $E_{\qqb}$ in the infrared region.

Such a behavior of $\est$ makes it more or less obvious that the moving string will have similar infrared properties as that at rest \cite{az3}. Indeed, if the effective tension has a local minimum at $r=\rmin$, then the string cannot get deeper than $\rmin$ in $r$ direction because a repulsive force prevents it from doing so. One can think of this as a soft wall located at $r = \rmin$. In this case, the large distance behavior of the string is completely determined by the wall. Concretely, a short inspection shows that $\ell(\r0)$ and $E_{\qqb}(\r0)$ are monotonically increasing in the interval $[0,\rmin]$. They become infinite at $\r0=\rmin$, where

\begin{equation}\label{large-l}
\ell(\r0)=-k\ln(\rmin -\r0)\,+O(1)
\,,\qquad
E_{\qqb}=-k\sigma\ln(\rmin -\r0)+O(1)
\,,
\end{equation}
with coefficients
\begin{equation}\label{sigma-ph}
k=2\sqrt{\est/f\est''}(\rmin)\,,
\qquad
\sigma=\g\est(\rmin)
\,.
\end{equation}
It follows from this that 

\begin{equation}\label{E-largel}
	E_{\qqb}=\sigma\ell+O(1)
	\,.
\end{equation}
Thus, this phase can indeed be interpreted as the phase of confinement. Note that $\sigma$ is the physical string tension.

On the other hand, if the effective string tension has no local minimum, then the string can get deeper into the bulk and finally reach the induced horizon. Its large distance behavior now is determined by the near horizon geometry. The crucial point here is that the effective string tension vanishes on the induced horizon. So it seems reasonable to expect that there is no linear term in the expression for $E_{\qqb}$ and therefore interpret this phase as the phase of deconfinement. However, the reality does not meet the expectation. The problem is that the connected configuration exists only if $\ell$ does not exceed the critical value $\lmax$.\footnote{This fact motivates the use of an operational definition of the screening length $\ell_s=\lmax$ \cite{uaw-wind}. However, it remains to be seen to what extent such a length will be consistent with the Debye screening length of lattice QCD \cite{lattice-rev}.} One way to address this problem is to introduce one more diagram as suggested in \cite{bak}. But we will not do so. Instead, we will discuss the Debye screening mass along the lines of \cite{a-screen}.

\begin{figure*}[htbp]
\centering
\includegraphics[width=6.3cm]{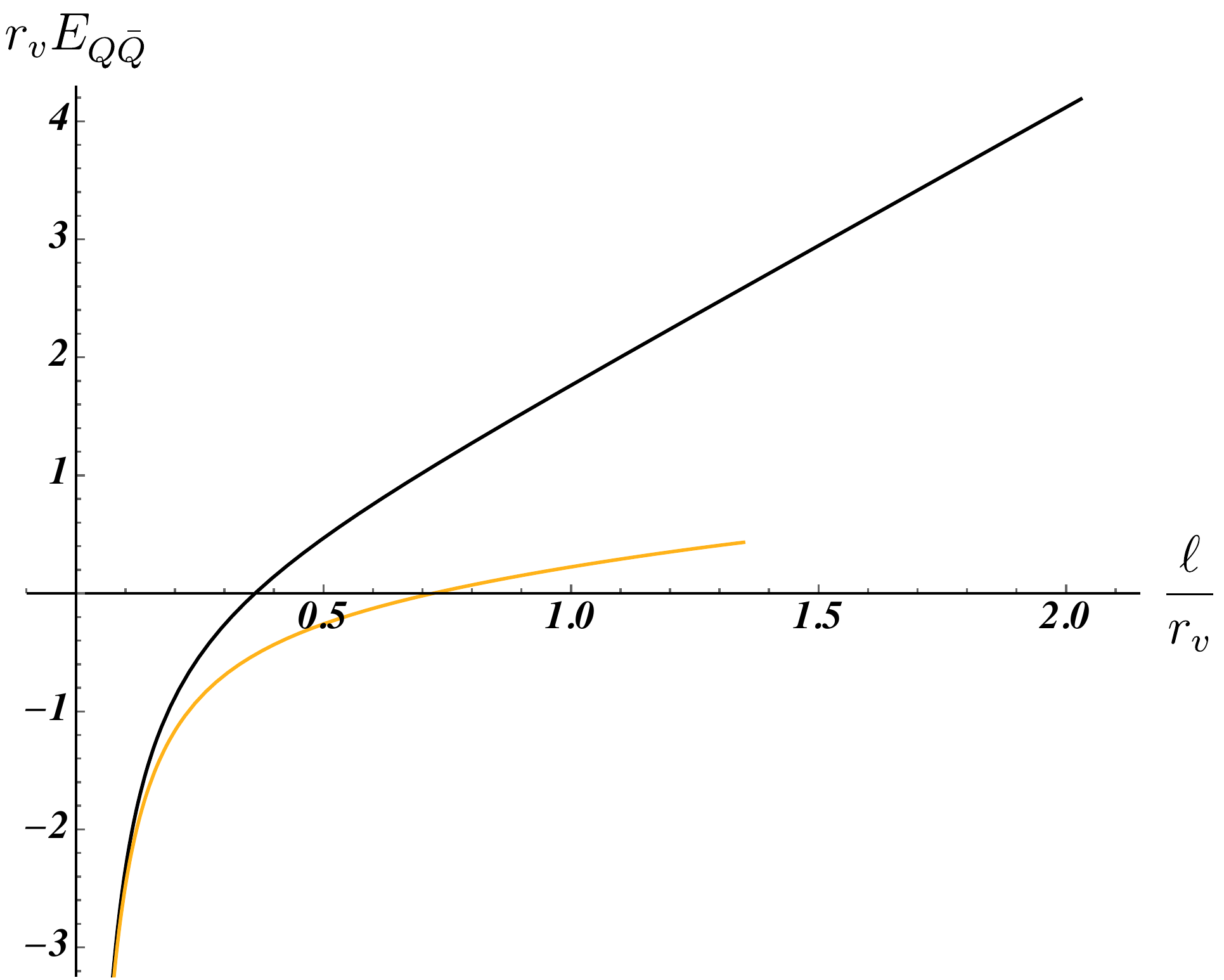}
\caption{{\small $E_{\qqb}$ versus $\ell$. The upper curve with $v=0.1$ and $\s\rv^2=1.3$ corresponds to the confined phase, while the lower with $v=0.2$ and $\s\rv^2=5.0$ to the deconfined phase. In both cases $\g=0.176$ and $c=0$.}}
\label{El}
\end{figure*}

We close this subsection with one example which illustrates the difference between the behavior of $E_{\qqb}$ in the confined and deconfined phases. In Figure \ref{El} we plot 
$E_{\qqb}$ as a function of $\ell$. Although we restrict ourselves here to the blackening factor \eqref{f-az2}, this picture holds for \eqref{f-noro} as well. 

\subsection{A disconnected string configuration}

Since a disconnected string configuration is so important in the deconfined phase, we will now study it. This configuration is similar to that of Figure \ref{configs} on the right, except that $\rh$ is replaced by $\rv$. It consists of two similar parts, each of which represents a single quark (antiquark) placed on the boundary and connected to the induced horizon by a Nambu-Goto string. 

We want, as in subsection A, to find the total energy of those strings which is twice larger than the energy of each string. The latter can be computed along the lines of \cite{a-pol}. Since the configuration in static, we choose the static gauge such that $t=\xi_1$ and $r=a\xi_2 +b$. From the boundary conditions $r(0)=0$ and $r(1)=\rv$, we have $a=\rv$ and $b=0$. The Nambu-Goto action is then

\begin{equation}\label{NG-Q}
S=-\g\int dt \int_0^{\rv} dr\,\est
\sqrt{x'{}^2+f^{-1}}
	\,,
\end{equation}
where a prime stands for a derivative with respect to $r$. It is obvious that $x(r)=const$ is a solution to the equation of motion that represents a straight string stretched between the boundary and the induced horizon. Moreover, this solution corresponds to the lowest string energy. 

In proceeding in this way, the important fact can be obscured by the gauge condition $r=a\xi_2+b$. So it is instructive to consider another, namely $x=a\xi_2+b$. Although this gauge condition is not convenient for studying the straight string, it is however helpful in understanding the force balance at the string endpoint located on the induced horizon. After a simple calculation, we find a boundary term in the variation of the Nambu-Goto action  

\begin{equation}\label{NG-b}
	\delta S_b=-\g\int dt\, \frac{\est \,r'}{\sqrt{1+f^{-1}r'{}^2}}\,\delta r\,\bigg\vert_0^{\rv}
\,.
\end{equation}
A prime now means a derivative with respect to $x$. The integrand provides the string tension in the $r$-direction.\footnote{By contrast, the string tension in the $x$-direction can easily be found by analyzing the action \eqref{NG-Q}.} For the straight string it is simply $\sigma_r=-g\est\sqrt{f}$. This tension vanishes on the induced horizon that implies that there is no need for an external force to keep the string endpoint in place. In other words, the configuration is in mechanical equilibrium.

The string energy is simply obtained by evaluating the Lagrangian in Eq.\eqref{NG-Q} on the solution. We have

\begin{equation}\label{EQ}
E_\qu=\g\int_0^{\rv}\frac{dr}{\sqrt{f}}\est
\,.
\end{equation}
The integral is is divergent at $r=0$. As before, we regularize it by imposing the cutoff $\epsilon$. Therefore, the regularized expression behaves for $\epsilon\rightarrow 0$ as

\begin{equation}\label{EQ1}
E_{\qu}^{\text{\tiny R}}=\frac{\g}{\epsilon}+E_{\qu}+O(\epsilon)
\,.
\end{equation}
After subtracting out the $\frac{1}{\epsilon}$ term and letting $\epsilon=0$, we get 

\begin{equation}\label{EQ2}
E_\qu=\g\int_0^{\rv}dr\biggl(\frac{\est}{\sqrt{f}}-\frac{1}{r^2}
\biggr)\,\,-\frac{\g}{\rv}+c
\,,
\end{equation}
with the same normalization constant as in \eqref{E2}. Clearly, $E_{\aqu}$ is also given by this expression. Note that for $v=0$, it reduces to the formula obtained in \cite{a-pol} for the heavy quark free energy. As such, $E_\qu$ can be interpreted as a heavy quark energy in the presence of a hot wind.


\section{More detail on the string tension and Debye screening mass}
\renewcommand{\theequation}{4.\arabic{equation}}
\setcounter{equation}{0}

Here we will describe more precisely important subtleties that arise when a heavy quark-antiquark pair moves though the thermal medium. Equivalently, these are related to the effects of the hot wind described in section III. 

We start with the simplest model in which the blackening factor is given by \eqref{f-az2}. In this case we can explicitly write down the expressions for the induced blackening factor 
and horizon 
\begin{equation}\label{frv}
\fv=1-\frac{r^4}{\rv^4}
\,,
\qquad
\rv=\rh\sqrt[4]{1-\vp^2}
\,.
\end{equation}
This allows a drastic simplification of the problem of determining local extrema of the function $\est (r)$. A simple argument using the properties of the cubic equation shows that it depends on a parameter $\s\rv^2$. The extrema exist only if $\s\rv^2 >\frac{\sqrt{27}}{2}$ and occur at 

\begin{equation}\label{roots}
\rmin=\rv\sqrt{\frac{2}{\sqrt{3}}\sin\phi}
\,,\qquad
\rmax=\rv\sqrt{\frac{2}{\sqrt{3}}\sin
\Bigl(\frac{\pi}{3}-\phi\Bigr)}
\,,
\end{equation}
where $\phi=\frac{1}{3}\arcsin\gamma\frac{T^2}{T_{pc}^2}$ and $T_{pc}=\frac{1}{\pi}\sqrt{\frac{2\s}{3\sqrt{3}}}$. Accordingly, the condition for confinement can now be written as 

\begin{equation}\label{confined}
\sqrt{\gamma}\,\frac{T}{T_{pc}}<1
\,.
\end{equation}
Thus $v$ goes to zero as $T$ approaches $T_{pc}$.

Now let us pause here for a moment to make a couple of remarks. First, $T_{pc}$ is a temperature which was originally introduced in \cite{az3}. Although it can be interpreted as a critical temperature of the model, this requires a caveat: in contrast to lattice QCD \cite{lattice-rev}, the string tension does not vanish at $T=T_{pc}$ (see the left panel of Figure \ref{sigma}). Second, it is unclear to us how to extend the above analysis to include the blackening factor \eqref{f-noro}. Therefore we will examine this case numerically. 
\subsection{The string tension}

Using the formula \eqref{sigma-ph} for the string tension and the expression \eqref{roots} for $\rmin$, we deduce a formula for the model with the blackening factor \eqref{f-az2}

\begin{equation}\label{tension-Tv}
\sigma(\vp,T)=\frac{1}{3}\sigma\gamma\frac{T^2}{T_{pc}^2}
\sqrt{\sin^{-2}\phi-\frac{4}{3}}
\,
\exp\biggl(3\gamma^{-1}\frac{T^2_{pc}}{T^2}\sin\phi -1\biggr)
\,,
\end{equation}
where $\sigma=\ep\g\s$ is the string tension at $T=\vp=0$ \cite{az1}. This result may be obtained in a simpler manner by rescaling $T$ by a factor of $\sqrt{\gamma}$ in the corresponding formula of \cite{az3}. It seems to be easy, but it is not without pitfalls, as we will see in the Appendix B. Notice that on the critical curve $\sqrt{\gamma}\frac{T}{T_{pc}}=1$ the tension is non-vanishing and approximately equal to $0.9\sigma$. 

 It is instructive to examine the low-temperature behavior of $\sigma(v,T)$. In view of Eq.\eqref{confined}, we expand it in powers of $\sqrt{\gamma}\frac{T}{T_{pc}}$. Then its expansion reads\footnote{It is worth mentioning that the leading low-temperature correction found within the four-dimensional string model \cite{pis} at zero velocity is proportional to $T^2$.}

\begin{equation}\label{app-tension}
\frac{\sigma(v,T)}{\sigma}\approxeq 1-\frac{2}{27}\gamma^2\frac{T^4}{T_{pc}^4}
\,.
\end{equation}
Here we drop the higher order terms. It turns out that this is a very good approximation for $\sqrt{\gamma}\frac{T}{T_{pc}}<0.6$. For larger values the discrepancy becomes clear and reaches a maximal value on the critical curve, where it is of order $3\%$.

\begin{figure*}[htbp]
\centering
\includegraphics[width=6.7cm]{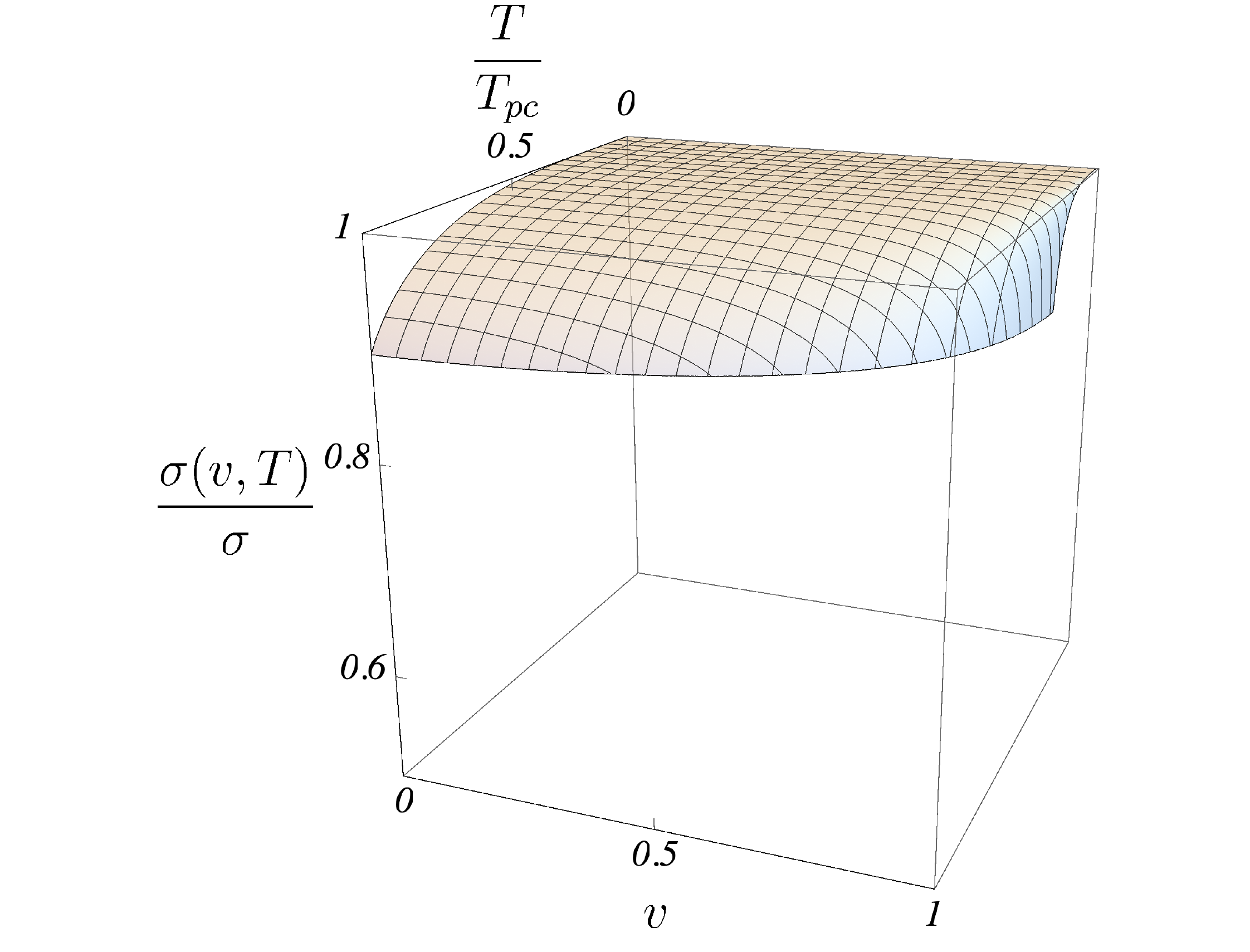}
\hspace{1.75cm}
\includegraphics[width=5.4cm]{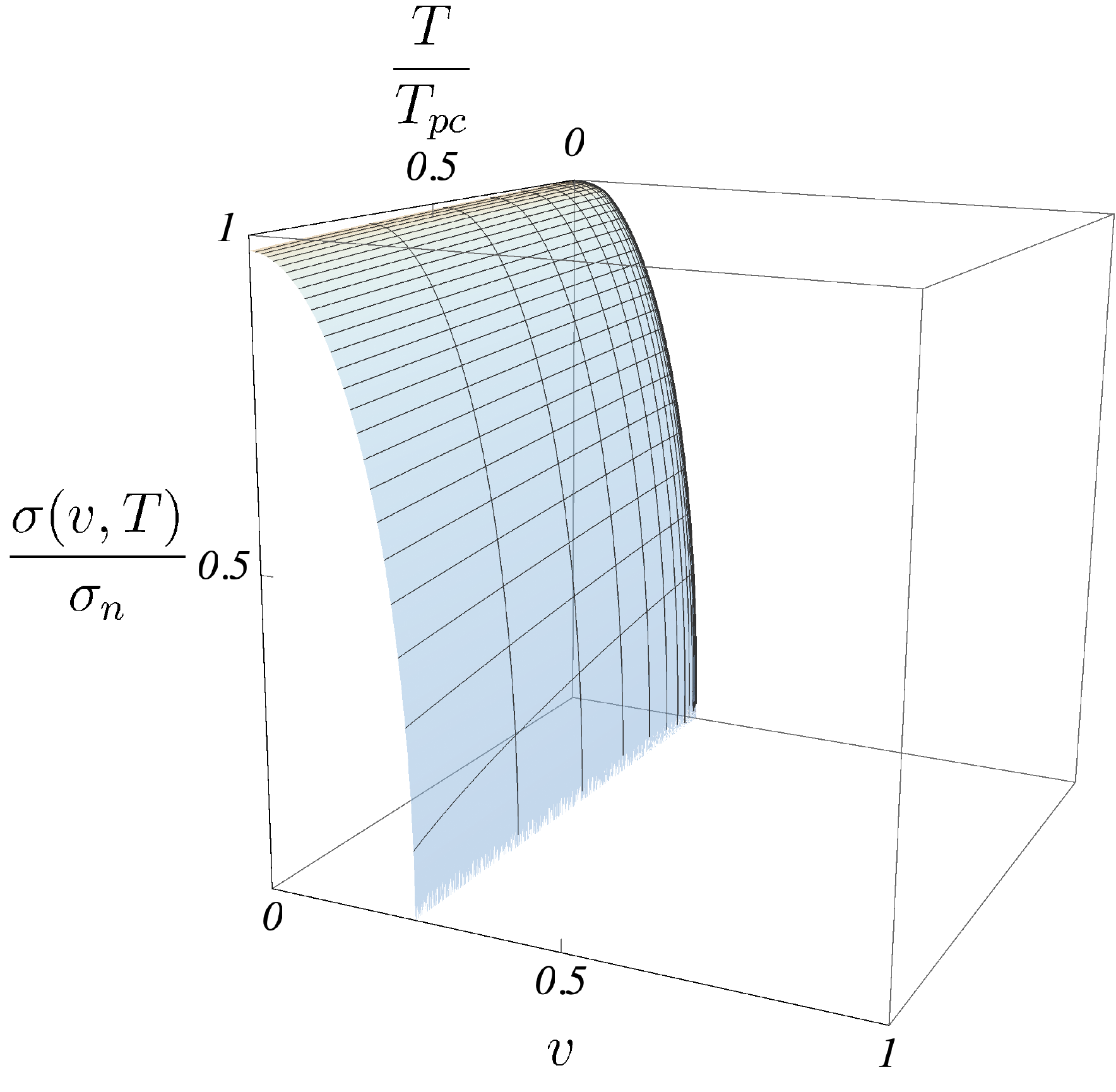}
\caption{{\small The string tension as a function of $v$ and $T$. The left and right panels represent the models with the blackening factors \eqref{f-az2} and \eqref{f-noro}, respectively.}}
\label{sigma}
\end{figure*}

To complete our discussion of this model we plot the string tension as a function of temperature and velocity. From the left panel of Figure \ref{sigma} we see that $\sigma(v,T)$ decreases with $T$ and $v$. Moreover, it is a slowly varying function except 
near the critical curve (especially for high velocities). For $T=0$  the tension is independent of $\vp$, as expected due to Lorentz invariance.

By contrast, algebra does not help much in understanding the model in which the blackening factor is given by \eqref{f-noro}. In that case, numerics is the only way. There are two special features to note in this example. The first is that $T_{pc}\approx 2\cdot 10^{-3}\sqrt{\s}$ which is two orders lower than that in \eqref{roots}.\footnote{For comparison, note that the estimate $T_c=\frac{\sqrt{\s}}{\pi}$ made in \cite{az2} for the critical temperature seems more in line with the lattice value given in \cite{boyd}.} The second is that $f$ does not go to $1$ as $T$ goes to $0$. In other words, this model does not reduce to that of \cite{az1} at $T=0$. Therefore, in practice, we can study this model only in a narrow temperature interval just above zero temperature. We take it from $4\cdot 10^{-7}T_{pc}$ to $T_{pc}$. For $v=0$, the string tension is approximately $\sigma_n=1.63\g\s$ at the lower limit. In the right panel of Figure \ref{sigma}, we plot the result for the string tension. We see that $\sigma(v,T)$ is a slowly varying function of temperature as before, but now it shows a steep decline as a function of velocity. In fact, the string tension vanishes near $v\approx 0.27$. We return to this in section V, after describing the Debye screening mass and gaining some information about its behavior in the temperature range close to $T_c$.

It is noteworthy that what we have discussed above is applicable only for confining backgrounds (confining gauge theories). For non-confining ones, including the AdS Schwarzschild black hole, it makes no sense as the string tension is vanishing. 

\subsection{The Debye screening mass}

By now we understand that at zero velocity the Debye screening mass can be computed from the correlator of two Polyakov loops \cite{lattice-rev}. Moreover, it can be done for both channels (singlet and octet) with the same result. Here there is an important issue that arises when one tries to split the correlator in two parts non-perturbatively.\footnote{This issue is still under debate \cite{octet}.} We won't discuss it here, in part because in string models the octet channel is not understood even on the very basic level, and in part because for our purposes it is enough, as shown before for $v=0$ \cite{a-screen}, to use the singlet channel.  

From here, we consider only the singlet channel. This means that the quark-antiquark pair is in the color singlet state. For this channel the string configuration is given by the connected configuration of section III. We define a binding energy of the pair as a difference between the energies of the connected and disconnected string configurations

\begin{equation}\label{bindE}
\Delta E_{\qqb}=E_{\qqb}-E_{\qu}-E_{\aqu}
\,,
\end{equation}
where $E_{\qqb}$ and $E_{\qu}(E_{\aqu})$ are given by Eqs.\eqref{E2} and \eqref{EQ2}, respectively. The above definition implies that the quark contributions, if any, cancel each other. In particular, this happens for the normalization coefficient $c$. Thus, the binding energy is expressed in terms of the string energies only. In the limit $v\rightarrow 0$, it reduces to $\Delta F_{\qqb}=F_{\qqb}-F_{\qu}-F_{\aqu}$, with $F$'s free energies. $\Delta F_{\qqb}$ plays a special role in determining the Debye screening mass via the singlet channel \cite{a-screen}. For more explanations, see the Appendix A.

Now we want to find the behavior of the binding energy at short distances. From Eqs.\eqref{El-small} and \eqref{EQ2}, we immediately obtain that 

\begin{equation}\label{B-Q}
\Delta E_{\qqb}=-\frac{\alpha_{\qqb}}{\ell}+
2\g\biggl[\frac{1}{\rv}+
\int_0^{\rv}dr\biggl(\frac{1}{r^2}-\frac{\est}{\sqrt{f}}\biggr)
\biggr]\,\,+o(1)
\,.	
\end{equation}
Motivated by the analysis in the Appendix A, we parameterize the binding energy as

\begin{equation}\label{E-gauge}
\Delta E_{\qqb}=-\frac{4\alpha}{3\ell}\ep^{-m\ell}+E_0
\,,
\end{equation}
where $E_0$ is an $\ell$-independent constant which reduces to that of Eq.\eqref{ident} as $v\rightarrow 0$. 

Expanding both sides of this equation in powers of $\ell$ and keeping only two leading terms in the expansion, at leading order we obtain $\alpha_{\qqb}=\frac{4}{3}\alpha$ which is the same relation as in the static case. But at the next order we have that

\begin{equation}\label{m-v}
m=\frac{\Gamma^{4}\bigl(\tfrac{1}{4}\bigr)}{4\pi^3}
\,\biggl[\frac{1}{\rv}+
\int_0^{\rv}dr\biggl(\frac{1}{r^2}-\frac{\est}{\sqrt{f}}\biggr)
+\mathbf{c}
\biggr]
\,,
\end{equation}
with $\mathbf{c}=-\frac{E_0}{2\g}$. Once we set the velocity to zero, the integral can be performed with the final result given by \eqref{m-v0}. 

One can think of $E_0$ as encoding information about quantum corrections to the classical energy of the string as well as about possible ambiguities in lattice QCD (at $v=0$). To this we have nothing to add, and will proceed with a couple of operational definitions for the screening mass. Consider the simplest possible forms of $\mathbf{c}$. On dimensional grounds, $\mathbf{c}$ may be proportional either to $T$ or to $1/\rv$. In the first case the simplest option is to take 

\begin{equation}\label{c-T}
\mathbf{c}=\mathfrak{w}T
\,,
\end{equation}
while in the second

\begin{equation}\label{c-rv}
\mathbf{c}=
\frac{\mathfrak{w}}{\rv}\cdot
\begin{cases}
\pi^{-1}\,\,& \text{for}\qquad\eqref{f-az2}\,,
\\
\rh T
 &\text{for}\qquad\eqref{f-noro}
\,.
\end{cases}	
\end{equation}
In both cases $\mathfrak{w}$ is a proportionality constant which coincides with that of Eq.\eqref{m-v0}. So, we treat it as the model parameter. 

In contrast to the static case, we are unfortunately unable to perform the integral in \eqref{m-v} analytically. But it is still possible to do so in certain limiting cases. For $f$ defined by \eqref{f-az2}, we take the limit $v\rightarrow 1$, getting 

\begin{equation}\label{m-v=1}
\sqrt[4]{1-v^2}\,\frac{m}{T}=C
\,,
\end{equation}
where the constant $C$ is given by 
\begin{equation}
C=\frac{\Gamma^2\bigl(\tfrac{1}{4}\bigr)}{\sqrt{2\pi}}
\cdot
\begin{cases}
1 &
\text{ for}\qquad\eqref{c-T}\,,
\\
1+\mathlarger{
\frac{\Gamma^2\bigl(\tfrac{1}{4}\bigr)}{(2\pi)^{\frac{3}{2}}}\frac{\mathfrak{w}}{\pi}}
 &\text{ for}\qquad\eqref{c-rv}
\,.
\end{cases}	
\end{equation}
For the blacking factor \eqref{f-noro}, the story is less simple. To get the behavior \eqref{m-v=1}, one has to take a double limit in which $v\rightarrow 1$ and $T\rightarrow \infty$. The latter condition implies that both $f$'s coincide. Notice that a similar formula arises in hot $N = 4$ supersymmetric QCD \cite{uaw-wind} without any restrictions on $v$ and $T$. 

\begin{figure*}[htbp]
\centering
\includegraphics[width=6.3cm, height=5.8cm ]{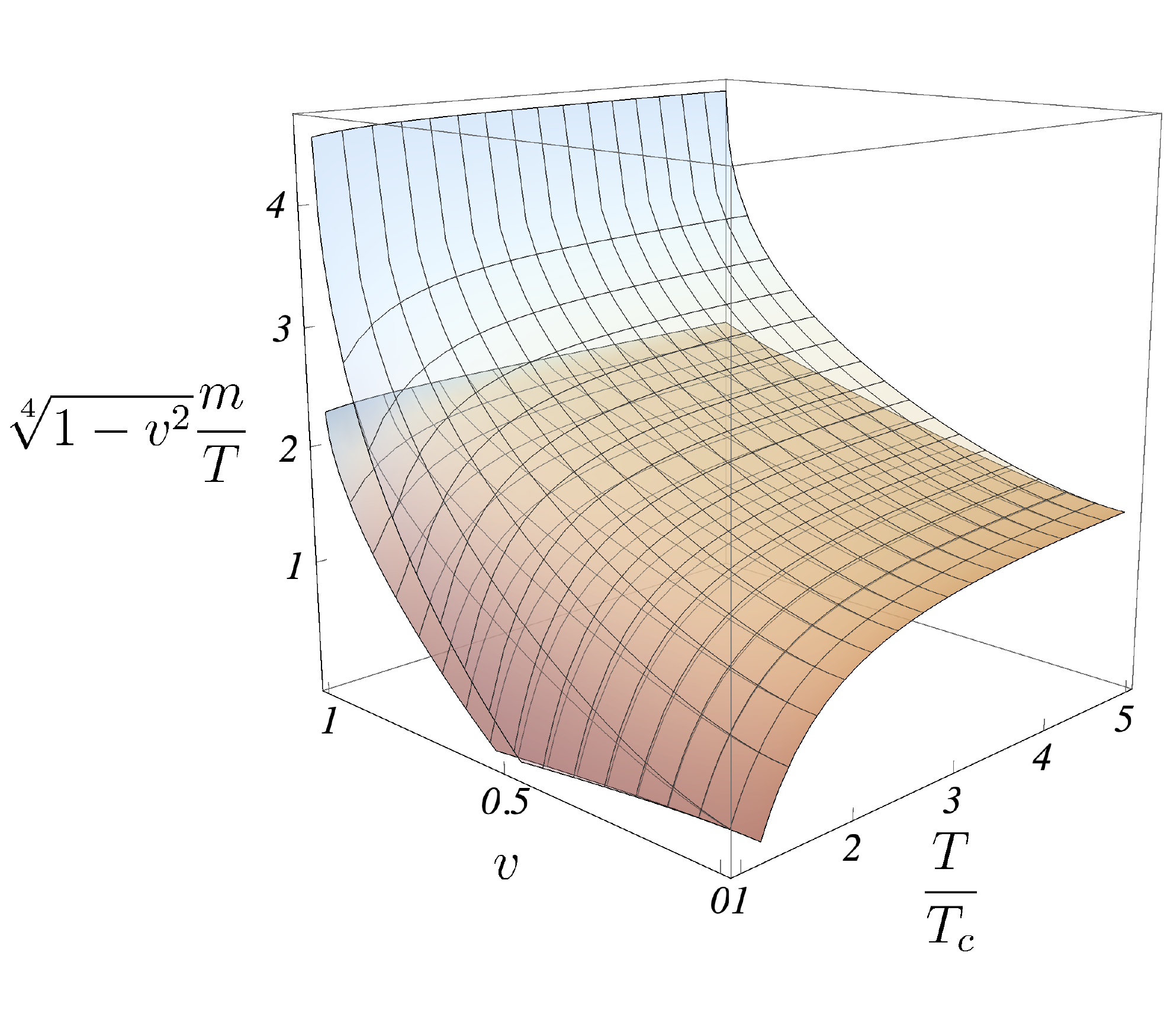}
\hspace{2.2cm}
\includegraphics[width=6.3cm, height=5.2cm]{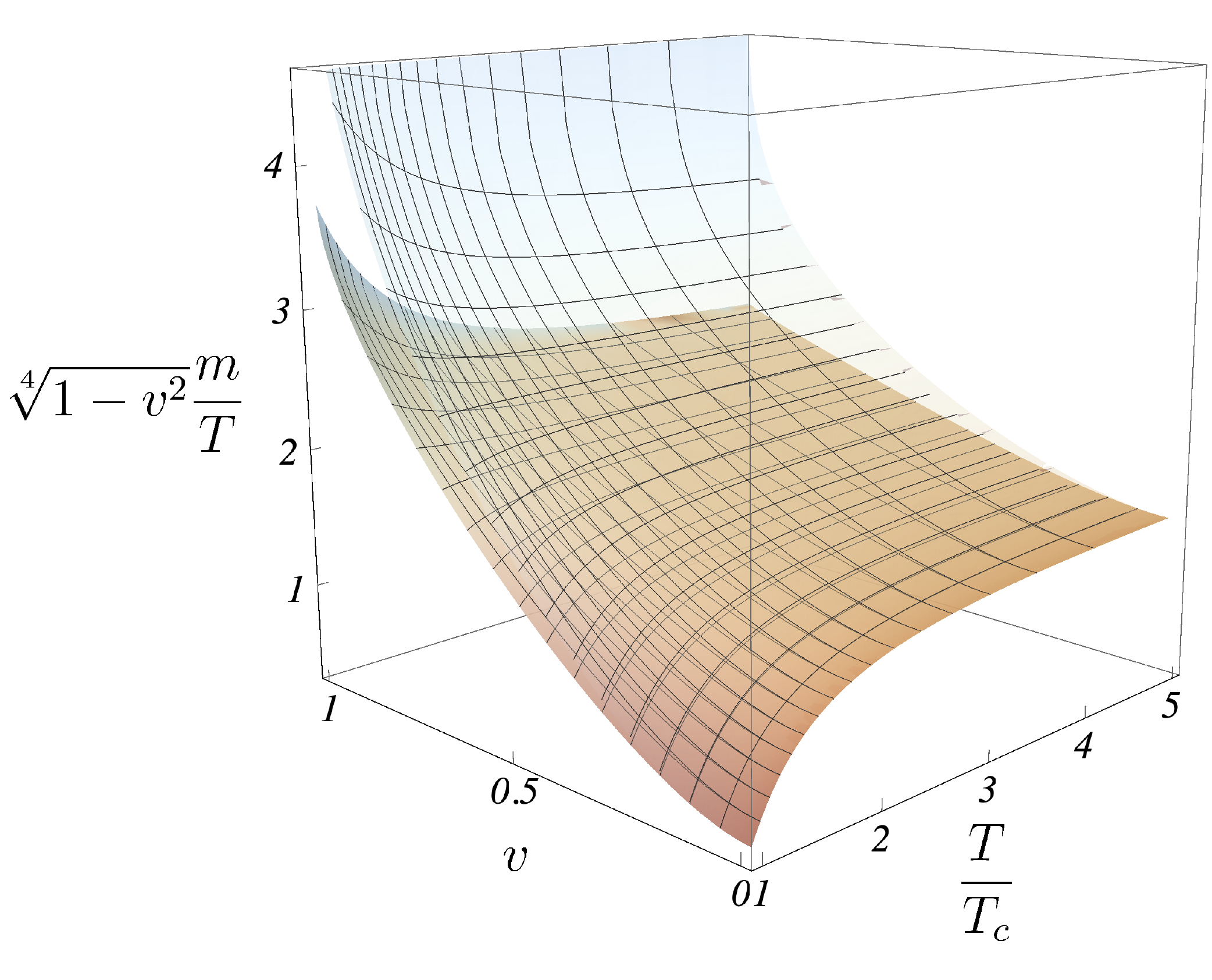}
\caption{{\small The Debye mass as a function of $v$ and $T$. The upper and lower surfaces correspond to \eqref{c-T} and \eqref{c-rv}, respectively. The left panel represents the model with the blackening factor \eqref{f-az2}, and the right with \eqref{f-noro}.}}
\label{mvT}
\end{figure*}

We go on further with a $SU(3)$ pure gauge theory, keeping in mind our goal of getting some intuition for phenomenology. Like in \cite{a-screen}, we write the critical temperature $T_c=0.629\sqrt{\sigma}$ \cite{boyd} as $T_c=0.435\sqrt{\s}$, where we used $\sigma=\ep\g\s$ and in the last step set $\g=0.176$ \cite{3q}. Then, the condition $T>T_c$ automatically implies that the models are in the deconfined phase as $T_c>T_{pc}$ in both cases. In Figure \ref{mvT} we plot the screening mass versus temperature and velocity. 
For none-zero velocity, it is convenient to normalize it also by a factor of $\sqrt{\gamma}$. \footnote{Thus, $\sqrt[4]{1-v^2}\frac{m}{T}\sim \frac{m}{m_{\text{AdS}}}$, as follows from \eqref{m-v=1}.} We set $\mathfrak{w}=-2.05$ to fit the lattice data at $v=0$ \cite{a-screen}. A common feature of these string models is that they fail to reproduce the lattice results in the near vicinity of the critical point as well as for high temperatures. Because of this, we take a small cutoff at $T=T_c$ and restrict $T$ to be less than $5T_c$. We see that the Debye screening mass is an increasing function of both variables, except for $v\gtrsim 0.72$ in the model with \eqref{f-noro}. It increases steeply near the edges $T=T_c$ and $v=1$. For non-relativistic velocities, the difference due to the two forms of $\mathbf{c}$ is negligible. As we will see in the next section, charmonium bound states may propagate through plasma only with low (non-relativistic) velocities, while bottomonium with $v\lesssim 0.82$. Therefore, the behavior of $m$ near $v=1$ could matter only for the analysis of topomonium bound states that seems hypothetical.
 
\section{Simple estimates}
\renewcommand{\theequation}{5.\arabic{equation}}
\setcounter{equation}{0}

Having understood how a wind velocity affects the string tension and Debye screening mass, we can formally redo the estimates of \cite{satz} for quarkonia moving through a thermal medium. In this section, we will do so in the context of finite temperature pure $SU(3)$ gauge theory. We assume that a relative motion of quarks is non-relativistic in the boosted (primed) frame.

We begin with an estimate of the characteristic size of quarkonium (ground state) in the confined phase at low temperature. Taking the quark-antiquark potential of the Cornell form $V(r)=-\frac{\alpha_{\qqb}}{r}+\sigma r +V_0$ and minimizing the ground state energy with respect to $r$, we see that it has a local minimum at 

\begin{equation}\label{rground}
\rho(v,T)=\Biggl
(\frac{1+\sqrt{1-\frac{4\mu^2}{27\sigma}\alpha^3_{\qqb}}}{2\mu\sigma}\,
\Biggr)^{\frac{1}{3}}
+
\frac{\alpha_{\qqb}}{3\sigma}
\Biggl(\frac{1+\sqrt{1-\frac{4\mu^2}{27\sigma}\alpha^3_{\qqb}}}{2\mu\sigma}\,
\Biggr)^{-\frac{1}{3}}
\,,
\end{equation}
with $\mu$ the reduced mass. The above expression is well defined for $\sigma\geq \frac{4\mu^2}{27}\alpha_{\qqb}^3$ that yields $\sigma\geq 9.67\times 10^{-4} \,\,\text{GeV}^2$ for charmonium states and $\sigma\geq 1.05\times 10^{-2}\,\,\text{GeV}^2$ for bottonium states. These values are obtained by setting $\alpha_{\qqb}=0.253$, as follows from \eqref{El-small} at $\g=0.176$, along with $m_c=1.27\,\text{GeV}$ and $m_b=4.18\,\text{GeV}$ \cite{pdg}. This means in practice that in the vicinity of the critical line the above estimate is not reliable.

\begin{figure*}[htbp]
\centering
\includegraphics[width=6.7cm, height=4.8cm]{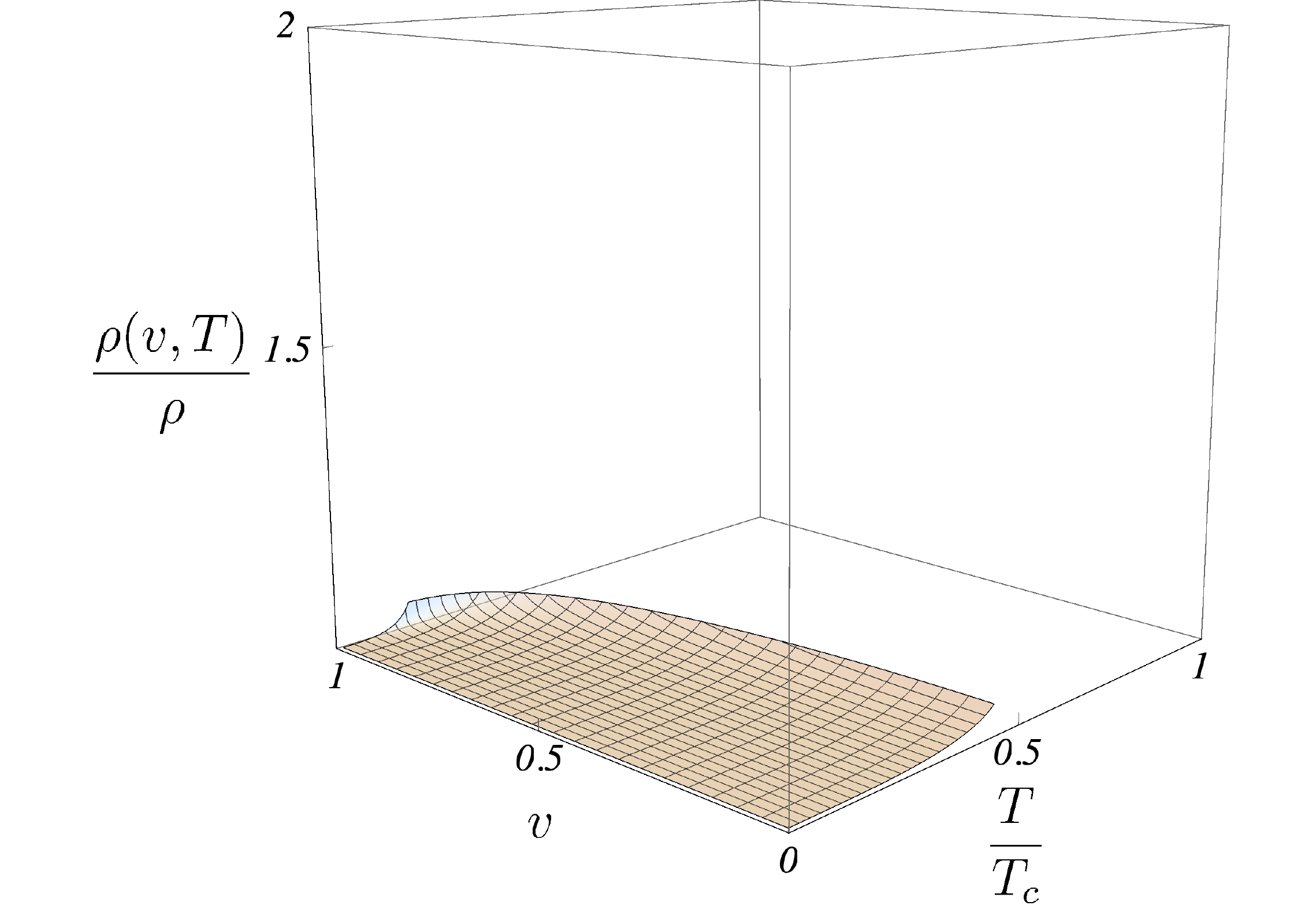}
\hspace{2.2cm}%
\includegraphics[width=5.8cm, height=4.9cm]{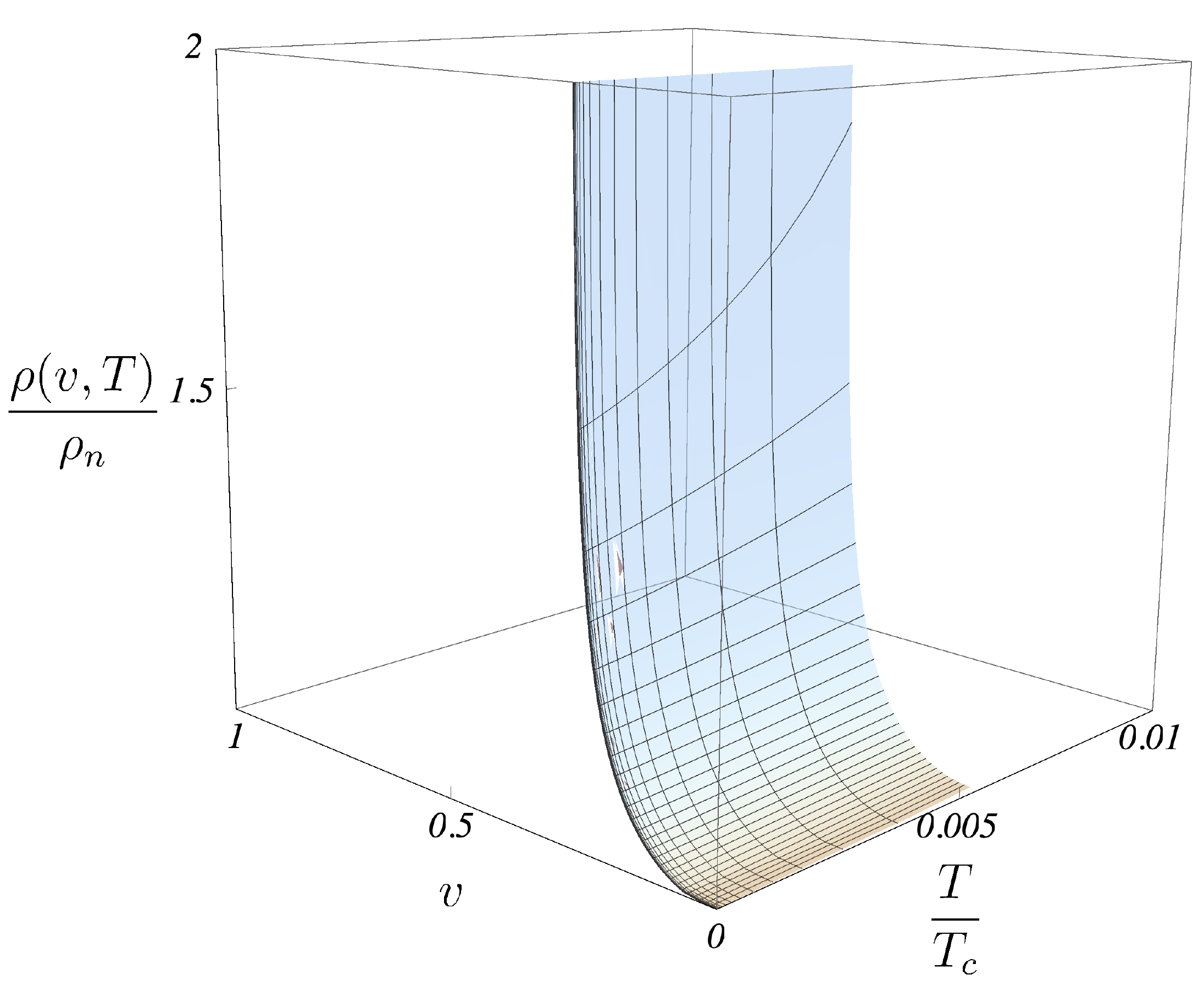}
\caption{{\small Typical graphs of $\rho(v,T)$. The left and right panels correspond to the models with the blackening factors \eqref{f-az2} and \eqref{f-noro}, respectively. We set $T_c=260\,\text{MeV}$, $\mu=2.09\,\text{GeV}$, $\rho=\rho(0,0)$, and $\rho_n=\rho(0, 5.2\cdot 10^{-7}\,\text{MeV})$.}}
\label{rho}
\end{figure*}

For illustrative purposes, in Figure \ref{rho} we show the plots of $\rho(v,T)$ for both forms of $\sigma(v,T)$ discussed in section IV. Here we  
assume that the string tension mildly depends on the wind direction. In the first case, when $f$ is given by \eqref{f-az2}, $\rho(v,T)$ can be found analytically by using the expression \eqref{tension-Tv}. The resulting formula is rather cumbersome to work with, so it seems natural to look for a simple approximation. We found that it can be well approximated by $1.25(\mu\sigma(v,T))^{-\frac{1}{3}}$. So, just like the string tension, the size of the ground state slowly varies with velocity and temperature, as also seen from the left panel of the figure. In the second case, with $f$ coming from \eqref{f-noro}, the restrictions of section IV remain and, in addition, it is now prohibited to approach close to the critical line where the approximation \eqref{rground} breaks down. We are unable to find a simple approximate formula for all velocities, but for low ones it is as described above. It is seen from the right panel that $\rho(v,T)$ increases rather slowly with temperature and steeply with velocity.

What we have learned above is that the characteristic size of quarkonium states increases with increasing wind velocity. So, it is natural to expect that quarkonium states will now dissociate at lower temperature. Our remaining goal will be to give another explanation of this and to make a simple estimate of the dissociation temperature.

First, let us recall one fact about the Debye screening mass. An effective parameterization for the lattice results is given by 

\begin{equation}\label{mv0}
\frac{m}{T}=a-b \frac{T_c^2}{T^2}
\,.
\end{equation}
It is motivated by the results obtained within the models of section II \cite{a-screen}. For $SU(3)$ pure gauge theory $a=0.71$ and $b=0.24$. Such a parameterization is quite good in the temperature range $1.06\, T_c\lesssim T\lesssim 3\,T_c$. 

In the deconfined phase, the analysis proceeds as before, but this time the heavy quark potential is taken to be of the form $V(r)=-\frac{\alpha_{\qqb}}{r}\ep^{-mr}+V_0$. It can be shown that a bound state exists if $\frac{m}{\mu\alpha_{\qqb}}\leq 0.84$ \cite{satz}. Using the above expression for $m$, the dissociation temperature at zero velocity is 

\begin{equation}\label{dis0}
\Td=\frac{T_c}{a}
\biggl
(0.42 \frac{\mu\alpha_{\qqb}}{T_c}+
\sqrt{ab+\Bigl(0.42 \frac{\mu\alpha_{\qqb}}{T_c}\Bigr)^2}\,
\biggr)
\,.
 \end{equation}
Let us make a simple estimate of $\Td$. With the same parameter values as quoted above, the dissociation temperature is then $\Td=1.05\, T_c$ ($c\bar c$), $\Td=1.37\,T_c$ ($c\bar b$), and $\Td=2.54 \,T_c$ ($b\bar b$). As expected, bottomonium states dissociate at higher temperature. 

The formula \eqref{mv0} for $m$ can be interpreted similarly to that for the pressure \cite{pis-fb}. In a temperature range between the critical temperature and some characteristic temperature of perturbative QCD, the ratio $\frac{m}{T}$ is given by a series in powers of $\frac{1}{T^2}$. In the absence of the $\frac{1}{T^2}$ corrections this formula reduces to $\frac{m}{T}=a$, which is also known in the context of $\text{AdS/CFT}$ \cite{uaw-book}. In this case the estimate gives $\Td=0.73T_c$ ($c\bar c$), $\Td=1.12T_c$ ($c\bar b$), and $\Td=2.41T_c$ ($b\bar b$). This indicates that a reasonable formula has to include not just a constant term but also corrections. Notice, however, that the bottonium is less sensitive (at least) to the leading correction because of its large mass. Indeed, it follows from \eqref{dis0} that $\Td=0.84\frac{\mu\alpha_{\qqb}}{a}$ as $\mu\rightarrow\infty$. 
 
The generalization to non-zero velocities does not change much of what we have said above as it follows from the analysis in the Appendix B. We start with a formula similar to Eq.\eqref{mv0} 
\begin{equation}\label{m/T}
	\frac{m}{T}= a(v)-b(v)\,\frac{T_c^2}{T^2}
	\,,
\end{equation}
but this time with velocity-dependent coefficients which reduce to those of \eqref{mv0} at $v=0$. 

Since the Debye mass increases with increasing velocity, it is natural to expect that the dissociation temperature will decrease with it. As an example of this, consider the simplest model, i.e. that with \eqref{f-az2}. In this case, the coefficients take the form 

\begin{equation}\label{abv}
	a(v)=m_0(v)
	\,,\qquad
	b(v)=b\,\frac{m_1(v)}{m_1(0)}
	\,,
\end{equation}
where the $m$'s are given by \eqref{m-faz2}. The value of $\mathfrak{w}$ is determined from the condition $a(0)=a$. With $a=0.71$, it is $\mathfrak{w}=-2.63$. A simple but crude approximation to \eqref{abv} is to assume that 

\begin{equation}\label{ab-uaw}
	a(v)=a\gamma^{\oh}
	\,,\qquad
	b(v)=b\gamma^{-\oh}
	\,.
\end{equation}
This follows from a rescaling argument (see the Appendix B). We have also assumed that the Debye screening mass mildly depends on the wind direction.

\begin{figure*}[htbp]
\centering
\includegraphics[width=7cm, height=5cm]{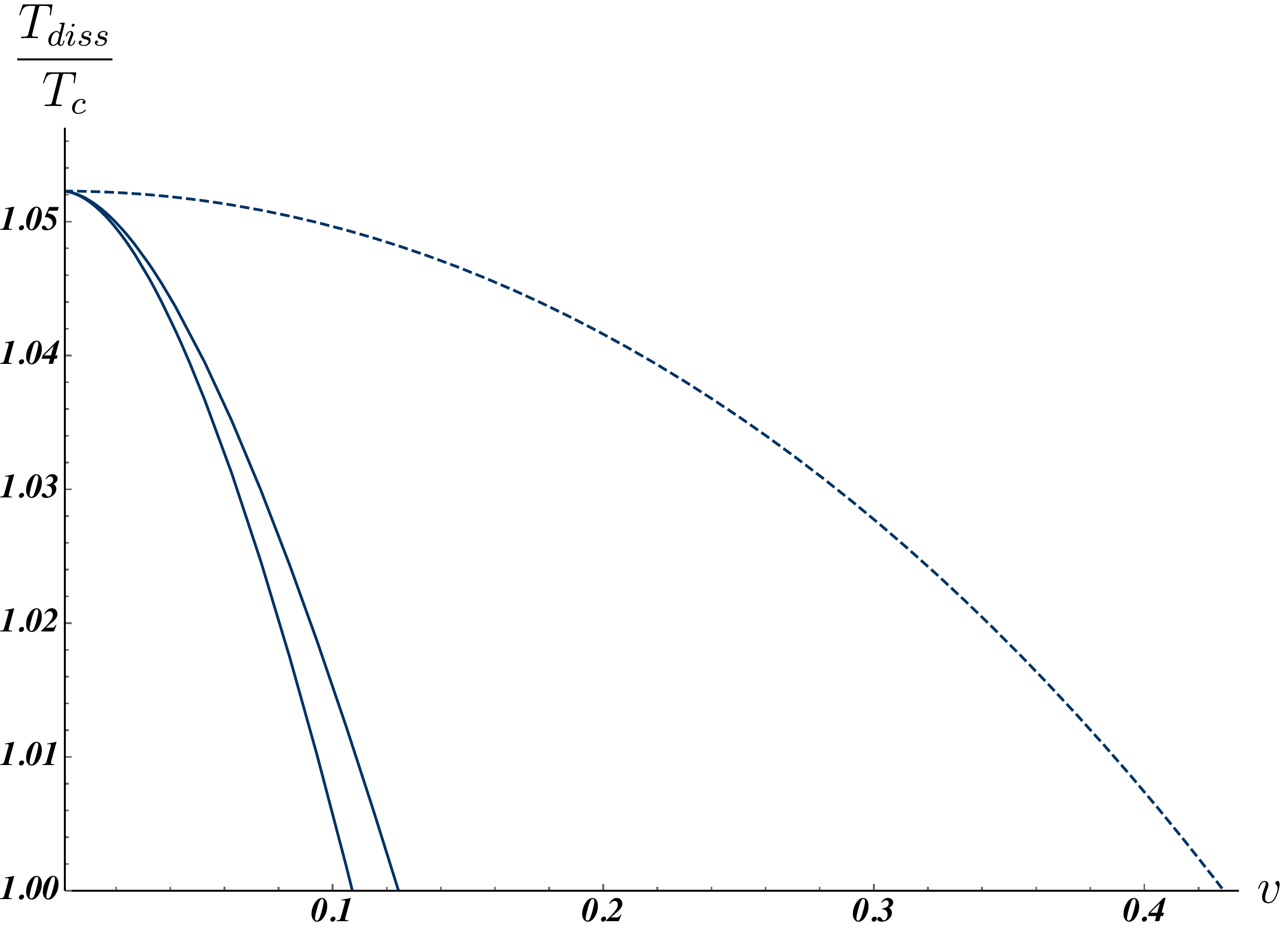}
\hspace{2.5cm}
\includegraphics[width=7cm, height=5cm]{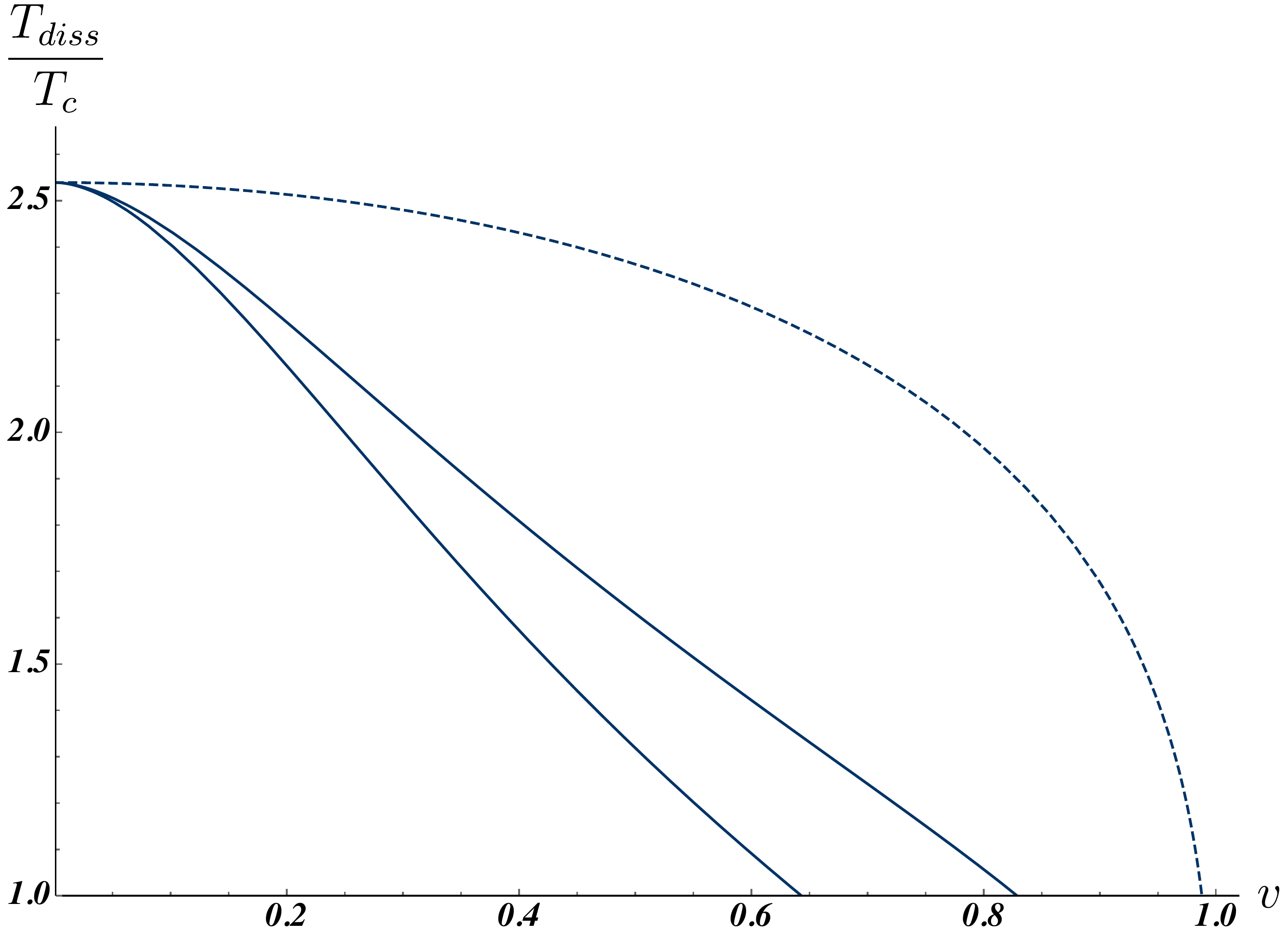}
\caption{{\small The dissociation temperature vs velocity, as shown on the left panel for charmonium states and on the right for bottomonium. The lower and upper solid curves correspond to \eqref{c-T} and \eqref{c-rv}, whereas the dashed curve is defined by \eqref{Tdis-uaw}, as in AdS/CFT.}}
\label{Tdis}
\end{figure*}

Now, using Eqs.\eqref{dis0}-\eqref{ab-uaw}, we can make a simple estimate of the dissociation temperature at finite velocity. In Figure \ref{Tdis}, we plot the results  
for charmonium and bottomonium states. We see that in the presence of a thermal wind  quarkonium states dissociate at lower temperature, as expected. In particular, charmonium states turn out to be more sensitive to the wind than bottomonium ones. This suggests that charmonium states may propagate through the plasma only with non-relativistic velocities. From the left panel, we obtain an upper bound $v\lesssim 0.13$. For these low velocities, the difference due to the two forms of $\mathbf{c}$ turns out to be unimportant.

Finally, let us note that the approximation \eqref{ab-uaw} results in a very simple formula for the dissociation temperature: 

\begin{equation}\label{Tdis-uaw}
\Td(v)=\sqrt[4]{1-v^2}\,\Td(0)
\,.
 \end{equation}
It is excellent in the context of AdS/CFT \cite{uaw-book}, but rather crude in comparison with ours (more phenomenologically motivated). 

\section{Concluding Comments}
\renewcommand{\theequation}{6.\arabic{equation}}
\setcounter{equation}{0}

The purpose of this paper has been to further develop the five-dimensional effective string models as a theoretical tool for studying strongly coupled gauge theories. In particular, we feel that it could be a correct way for dealing with situations in QCD where evolution in real time is an essential feature. This is an optimistic view. Of course, there are many open questions left which deserve further study, and the main question: what is the string dual to QCD?

 Of particular interest for heavy-ion phenomenology is those backgrounds which describe the medium not only at finite temperature but also at finite baryon density. Furthermore, there is an important aspect related to the anisotropic expansion of the quark-gluon plasma. We believe that these and similar issues are worthy of future study.

\begin{acknowledgments}
We would like to thank I. Aref'eva, P. de Forcrand, R.R. Metsaev, P. Weisz, and U.A. Wiedemann for useful and encouraging discussions. We also thank the Arnold Sommerfeld Center for Theoretical Physics and CERN Theory Division for the warm hospitality. This work was supported in part by RFBR Grant 18-02-40069.
\end{acknowledgments}

\vspace{.35cm} 
\appendix
\section{The Debye screening mass}
\renewcommand{\theequation}{A.\arabic{equation}}
\setcounter{equation}{0}
One of the features of the quark-gluon plasma is the phenomenon of Debye screening. Its numerical characteristic is a screening mass which provides the scale over which interactions are effective. Here we briefly recall how the five-dimensional string models can be used to estimate the Debye mass along the lines of lattice QCD \cite{a-screen}.

In lattice QCD, the Debye mass can be determined from the exponential fall-off of correlation functions at asymptotically large separation. The simplest possible correlator is that of two Polyakov loops \cite{lattice-rev}. Since a quark-antiquark pair could be either in a singlet or in an octet state, it seems natural to write the correlator as a sum of the singlet and octet contributions 

\begin{equation}\label{LL-lat0}
\frac{\langle\, L^\dagger(x_1) L(x_2)\,\rangle}{\vert\langle\, L\,\rangle\vert^2}=\frac{1}{9}\ep^{-\tfrac{F^1_{\qqb}}{T}}+\frac{8}{9}\ep^{-\tfrac{F^8_{\qqb}}{T}}
\,.
\end{equation} 
Here $F^1_{\qqb}$ and $F^8_{\qqb}$ are the singlet and octet free energies, respectively. 

One parameterization of the $F$'s, motivated by high temperature perturbation theory, is simply

\begin{equation}\label{F1}
F^1_{\qqb}(\ell)=-\frac{4\alpha}{3\ell}\ep^{-m\ell}
\,,\qquad
F^8_{\qqb}(\ell)=\frac{\alpha}{6\ell}\ep^{-m\ell}
\,,	
\end{equation}
where $\ell=\vert x_1-x_2\vert$. Obviously, for small $\ell$ the singlet channel dominates and therefore determines the correlator 

 \begin{equation}\label{LL-lat}
\frac{\langle\, L^\dagger(x_1) L(x_2)\,\rangle}{\vert\langle\, L\,\rangle\vert^2}=\frac{1}{9}\ep^{-\tfrac{F^1_{\qqb}}{T}}
\,,\qquad\text{with}\qquad
F^1_{\qqb}(\ell)=-\frac{4\alpha}{3\ell}+\frac{4}{3}\alpha m+o(1)
\,.
\end{equation}

\begin{figure*}[htbp]
\centering
\includegraphics[width=7.15cm]{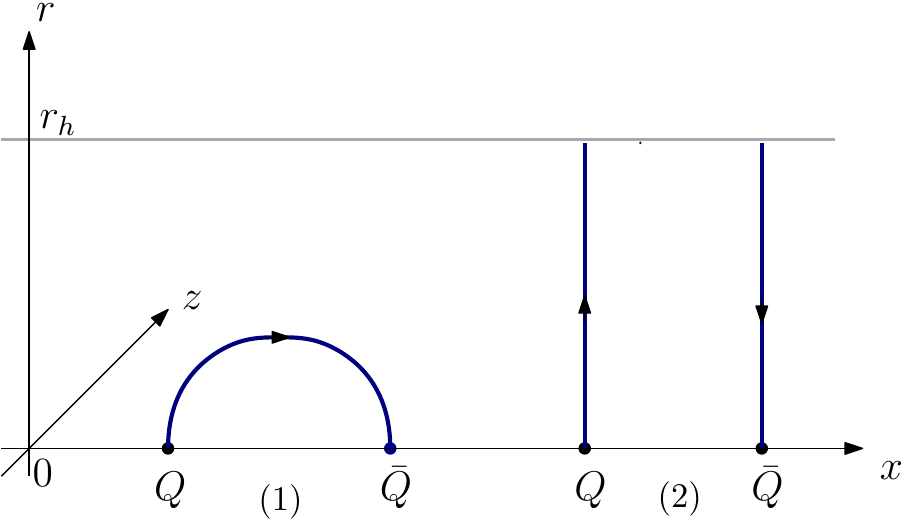}
\caption{{\small String configurations for the correlator at short distances. The horizontal line at $r=\rh$ represents the horizon.}}
\label{configs}
\end{figure*}

In discussing the correlator within five-dimensional effective models, one has to evaluate a worldsheet path integral obeying the boundary condition that the string worldsheet has the loops for its boundary. In practice, the integral can only be evaluated semiclassically in terms of minimal surfaces. The result is written as 

\begin{equation}\label{LL-string0}
\langle\, L^\dagger(x_1) L(x_2)\,\rangle=\sum_n w_n\,\ep^{-S^{(n)}}
\,.
\end{equation}
Here $S^{(n)}$ is a Nambu-Goto action evaluated on a classical solution (string configuration) whose relative weight is $w_n$. For small $\ell$, the configuration depicted in the left panel of Figure \ref{configs} dominates. 
In this case $S^{(1)}$ behaves as $S^{(1)}=\tfrac{1}{T}\bigl(-\frac{\alpha_{\qqb}}{\ell}+2c+o(1)\bigr)$, with a universal Coulomb coefficient.\footnote{This statement can be expressed as the space becomes asymptotically AdS near its boundary.} Such a behavior allows one, after fitting the Coulomb coefficients, to associate configuration (1) with the singlet channel. The normalization is provided by configuration (2) shown in the right panel. So, we have
 
 \begin{equation}\label{LL-string}
\frac{\langle\, L^\dagger(x_1) L(x_2)\,\rangle}{\vert\langle\, L\,\rangle\vert^2}=
\omega\,\ep^{-\tfrac{\Delta F_{\qqb}}{T}}
\,.
\end{equation}
Here $\Delta F_{\qqb}=F_{\qqb}-F_{\qu}-F_{\aqu}$, with $F_{\qqb}=TS^{(1)}$ and $F_{\qu}=F_{\aqu}=TS^{(2)}$. The factor $\omega$ is a ratio of the weights.

Equating the expressions \eqref{LL-lat} and \eqref{LL-string}, we get 

\begin{equation}\label{ident}
F^1_{\qqb}=\Delta F_{\qqb}-T\ln(9\omega)
 \,.	
\end{equation}
If we expand both sides of this equation in powers of $\ell$ and equate coefficients, we find $\alpha_{\qqb}=\frac{4}{3}\alpha$ at leading order, while at the next order we have $\alpha_{\qqb}m=-2F_{\qu}-T\ln(9\omega)$. Taking $w=\tfrac{\ep^{\s r^2}}{r^2}$ for the warp factor and using the explicit expression for $F_{\qu}$, we arrive at

\begin{equation}\label{m-v0}
m=\frac{\Gamma^{4}\bigl(\tfrac{1}{4}\bigr)}{4\pi^3}
\biggl(
\frac{\ep^h}{\rh}-\sqrt{\pi\s}\,\erfi\bigl(\sqrt{h}\,\bigr)+\mathfrak{w}T
\biggr)
\,,
\end{equation}
with $\mathfrak{w}=-\ln(9\omega)/2\g$. Note that the screening mass does not depend on the actions for the heavy quark sources, and in particular on the normalization constant $c$. For our purposes, we think of $m$ as a function of $T$, and treat $\mathfrak{w}$ as a model parameter. The reasoning for this is that at the moment a detailed calculation of the weight factors is beyond our abilities.

\section{Taylor Series of $m$}
\renewcommand{\theequation}{B.\arabic{equation}}
\setcounter{equation}{0}
Our goal here is to compute the first few coefficients of the Taylor series of $\frac{m}{T}(h)$ about $h=0$. This provides the basis for making the estimate of the dissociation temperature in Section V. The higher order coefficients can be computed in a similar way.

For both forms of $f$ and $\mathbf{c}$ we see explicitly that $\frac{m}{T}$ is regular at $h=0$. So we expand it in powers of $h$ 

\begin{equation}\label{m/T-Taylor}
	\frac{m}{T}(h)=m_0-m_1 h+O(h^2)
	\,.
	\end{equation}
This formula is completely equivalent to our earlier formula \eqref{mv0}, with $\frac{1}{T^2}$ expansion.

First, we consider the model whose blackening factor is of the form \eqref{f-az2}. After expending it in powers of $h$ and performing the corresponding integrals over $r$, we get

\begin{equation}\label{m-faz2}
m_0=
\frac{\Gamma^4\bigl(\tfrac{1}{4}\bigr)}{(2\pi)^2}
\gamma^{\oh}
\biggl(
C_{\text{\tiny A}}
+\oh v^2 B\bigl(\tfrac{3}{4},\tfrac{1}{2}\bigr)
F\bigl(\tfrac{3}{2};\tfrac{3}{4},\tfrac{5}{4};\gamma^{-2} \bigr)
\biggr)
\,,
\qquad
m_{1}=\frac{\Gamma^6\bigl(\tfrac{1}{4}\bigr)}{6(2\pi)^{\frac{5}{2}}}
\gamma^{-\oh}F\bigl(\tfrac{1}{2};\tfrac{1}{4},\tfrac{7}{4};\gamma^{-2} \bigr)
\,,
\end{equation}
with
\begin{equation}\label{CA}
C_{\text{\tiny A}}(v)=
\frac{\mathfrak{w}}{\pi}\cdot
\begin{cases}
\gamma^{-\oh} & \text{for}\quad\eqref{c-T}
\,,\\
1 &
\text{for}\quad\eqref{c-rv}
\,.
\end{cases}	
\end{equation}
Here $F(a; b, c; x)$ and $B(a,b)$ are the hypergeometric and beta functions \cite{gr}.\footnote{As usual, the subscripts in ${}_2F_1$ are omitted.} In the limit $v\rightarrow 1$ the coefficients become

\begin{equation}\label{m-faz21}
m_0=C\gamma^{\oh}
\,,
\qquad
m_{1}=0
\,,
\end{equation}
as expected from \eqref{m-v=1}. Here we used $F(a; b, c; 0)=1$ and $\oh B\bigl(\tfrac{3}{4},\tfrac{1}{2}\bigr)=\tfrac{(2\pi)^{\frac{3}{2}}}{\Gamma^2(\frac{1}{4})}$.

The $v$-dependence of these coefficients merits some comment. At first sight, it may seem that it is completely determined by Eqs.\eqref{Taz2} and \eqref{frv} so that the result may be obtained from that of \cite{a-screen} by rescaling $T\rightarrow \sqrt{\gamma}T$. This is partially true, but substantially false as seen from the explicit expressions of the coefficients. The reasons for this are as follows. First, the form of the metric \eqref{metricv} is not simple in sense that it includes not only $\fv$, but also $f$ and $v$. While the former does lead to the rescaling, the others do not. Second, the uncertainty related to the form of $\mathbf{c}$ may cause, as seen from \eqref{CA}.  

For the model defined by \eqref{f-noro}, the leading coefficient is given by \eqref{m-faz2}, while the next to leading is 

\begin{equation}\label{m-fnoro}
m_{1}=
\frac{\Gamma^4\bigl(\tfrac{1}{4}\bigr)}{16\pi^2}
\gamma^{-\oh}
\biggl(
\gamma C_{\text{\tiny B}}
+
B\bigl(\tfrac{1}{4},\tfrac{3}{2}\bigr)
\,F\bigl(\tfrac{1}{2};\tfrac{1}{4},\tfrac{7}{4};\gamma^{-2} \bigr)
+\frac{v^2}{2}B\bigl(\tfrac{5}{4},\tfrac{1}{2}\bigr)
\,F\bigl(\tfrac{3}{2};\tfrac{5}{4},\tfrac{7}{4};\gamma^{-2} \bigr)
-\frac{3}{2}v^2\gamma
B\bigl(\tfrac{3}{4},\tfrac{1}{2}\bigr)
\,F\bigl(\tfrac{3}{2};\tfrac{3}{4},\tfrac{5}{4};\gamma^{-2} \bigr)
\biggr)
\,,
\end{equation}
with
\begin{equation}\label{CB}
C_{\text{\tiny B}}(v)=\frac{\mathfrak{w}}{\pi}\cdot
\begin{cases}
0 & \text{for}\quad\eqref{c-T}
\,,\\
\gamma^{-1}-1 & \text{for}\quad\eqref{c-rv}
\,.
\end{cases}	
\end{equation}
In this form we can easily take the limit $v\rightarrow 1$ so that we find  

\begin{equation}\label{m1-noro}
m_{1}=
\frac{\Gamma^4\bigl(\tfrac{1}{4}\bigr)}{16\pi^2}
\gamma^{\oh}
\biggl(C_{\text{\tiny B}}(1)-\frac{3}{2}B\bigl(\tfrac{3}{4},\tfrac{1}{2}\bigr)\biggr)
\,.
\end{equation}
Thus, if $\mathfrak{w}=-2.05$, then $m_1$ is negative. This implies that for relativistic velocities the ratio $\frac{m}{T}$ decreases with the increase of temperature, as also seen from Figure \ref{mvT}. 

\begin{figure*}[htbp]
\centering
\includegraphics[width=6.5cm, height=5cm]{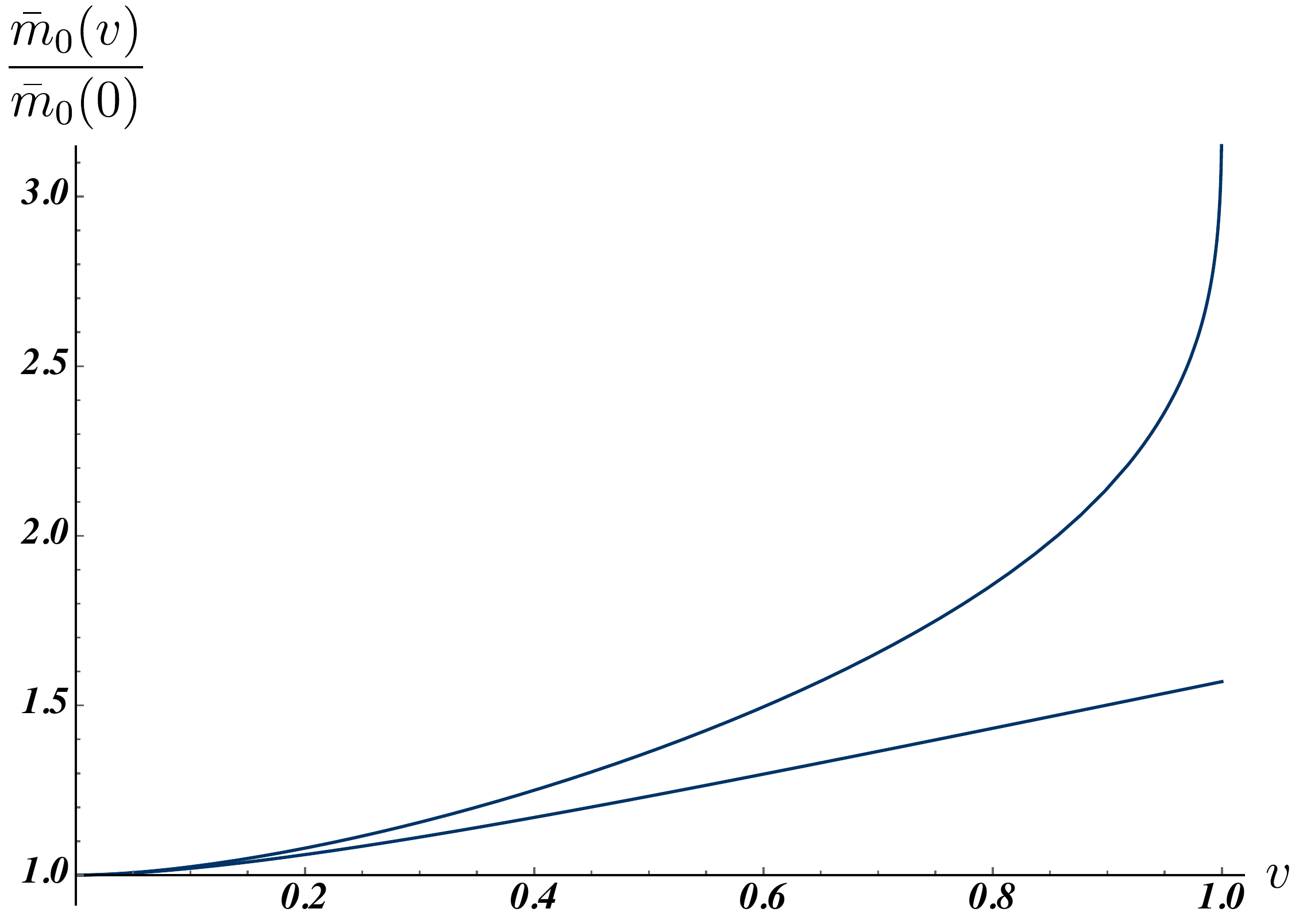}
\hspace{2cm}
\includegraphics[width=6.5cm, height=5cm]{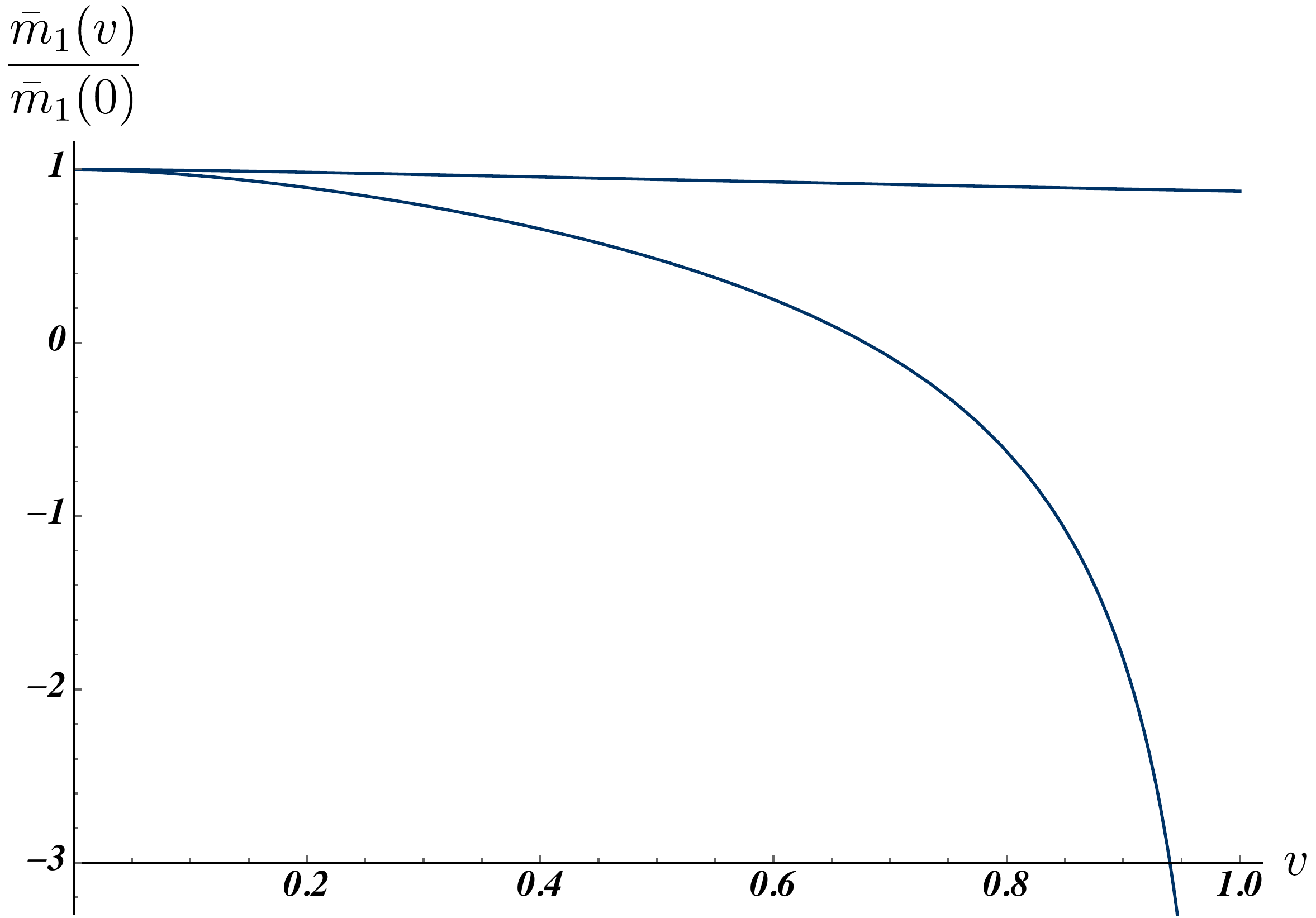}
\caption{{\small Test functions. Left: $\bar m_0$. The upper and lower curves correspond to \eqref{c-T} and \eqref{c-rv}, respectively. Here $\mathfrak{w}=-2.05$. Right: $\bar m_1(v)$. The upper curve is defined by \eqref{m-faz2}, whereas the lower by \eqref{m-fnoro} along with \eqref{c-T}.}}
\label{m01}
\end{figure*}

We conclude this appendix with some further remarks on the $v$-dependence of the coefficients. To understand deviation from the scaling $T\rightarrow \sqrt{\gamma}T$, it is convenient to introduce two test functions 

\begin{equation}\label{testm}
\bar m_0(v)=m_0\gamma^{-\oh}
\,,\qquad
\bar m_1(v)=m_1\gamma^{\oh}
\,.
\end{equation}
In Figure \ref{m01} we plot these functions normalized to their values at $v=0$.  
For both $C_{\text{\tiny A}}$, $\bar m_0$ is regular in the interval. It increases more rapidly for $C_{\text{\tiny A}}$ defined by \eqref{c-T}. By contrast, $\bar m_1$ is regular for \eqref{m-faz2} but singular for \eqref{m-fnoro}, where $m_1\sim\gamma$ as follows from the asymptotic formula \eqref{m1-noro}. It decreases with $v$, although very slowly for \eqref{m-faz2}. Thus, we see that treating both functions as if they were constants may be a rather crude approximation. 



\end{document}